\documentclass{article}
\usepackage{enumerate,amsmath,amssymb,amsfonts,amsthm,version}
\def\pmb#1{\setbox0=\hbox{$#1$}%
  \kern-.025em\copy0\kern-\wd0
  \kern.05em\copy0\kern-\wd0
  \kern-.025em\raise.0433em\box0}
\def\pmbs#1{\setbox0=\hbox{$\scriptstyle #1$}%
  \kern-.0175em\copy0\kern-\wd0
  \kern.035em\copy0\kern-\wd0
  \kern-.0175em\raise.0303em\box0}
\def\be{\begin{equation}}
\def\ee{\end{equation}}
\def\bea{\begin{eqnarray}}
\def\eea{\end{eqnarray}}
\def\lb{\label}
\def\ct{\cite}
\def\bi{\bibitem}
\def\vec#1{\mbox{\boldmath$#1$}}
\def\ten#1{\mbox{\boldmath$#1$}}
\def\mat#1{\mbox{\boldmath$#1$}}
\def\lgth{[\,\mbox{length}\,]}

\def\gam{\gamma}
\def\Gam{\Gamma}
\def\d{\delta}
\def\eps{\epsilon}
\def\th{\theta}

\def\sig{\sigma}

\def\Sig{\Sigma}

\def\om{\omega}
\def\Om{\Omega}
\def\Oml{\Omega_{\Lambda}}

\def\udot{\dot{u}}
\def\Udot{\dot{U}}
\def\cv{{\cal V}}

\def\cn{{\cal N}}

\def\ck{\Omega_{k}}

\def\D{\mbox{D}}
\def\vece{\vec{e}}
\def\vecu{\vec{u}}

\def\ptl{\partial}
\def\parb{\pmb{\partial}}
\def\e{{\rm e}}
\def\la{\langle}
\def\ra{\rangle}
\def\ti{\tilde}
\def\bigO{{\cal O}}
\def\Eta{\eta}
\def\hsp5{\hspace{5mm}}
\def\ann{{Ann.~Phys.~(N.Y.)} }
\def\apj{{Astrophys.~J.} }
\def\cmp{{Commun.~Math.~Phys.} }
\def\cqg{{Class.~Quantum~Grav.} }
\def\grg{{Gen.~Rel.~Grav.} }
\def\jmp{{J.~Math.~Phys.} }
\def\mn{{Mon.~Not.~R.~Astron.~Soc.} }
\def\prd{{Phys.~Rev.~D} }

\def\prs{{Proc.~R.~Soc.~Lond.~A} }
\newcommand{\enl}{\\\hfill\rule{0pt}{0pt}}
\begin{document}
%%%%%%%%%%%%%%%%%%%%%%%%%%%%%%%%%%%%%%%%%%%%%%%%%%%%%%%%%%%%%%%%%%%
\title{{\sc Asymptotic isotropization}\\
{\sc in inhomogeneous cosmology}}
%%%%%%%%%%%%%%%%%%%%%%%%%%%%%%%%%%%%%%%%%%%%%%%%%%%%%%%%%%%%%%%%%%%
\author{{\sc Woei Chet Lim$^{1}$\thanks{Electronic address:
{\tt wclim@math.uwaterloo.ca}}\ ,
Henk van Elst$^{2}$\thanks{Electronic address: {\tt
H.van.Elst@qmul.ac.uk}}\ ,
Claes Uggla$^{3}$\thanks{Electronic address:
{\tt Claes.Uggla@kau.se}}\ , %}\ , \\
John Wainwright$^{1}$\thanks{Electronic address:
{\tt jwainwri@math.uwaterloo.ca}}}\\
$^{1}${\small\em Department of Applied Mathematics, University of
Waterloo, Waterloo, Ontario, Canada N2L 3G1}\\
$^{2}${\small\em Astronomy Unit, Queen Mary, University of London,
Mile End Road, London E1 4NS, United Kingdom}\\
$^{3}${\small\em Department of Physics, University of Karlstad,
S-651 88 Karlstad, Sweden}}

%%%%%%%%%%%%%%%%%%%%%%%%%%%%%%%%%%%%%%%%%%%%%%%%%%%%%%%%%%%%%%%%%%%
\date{\normalsize{June 26, 2003}}
%%%%%%%%%%%%%%%%%%%%%%%%%%%%%%%%%%%%%%%%%%%%%%%%%%%%%%%%%%%%%%%%%%%
\maketitle
%%%%%%%%%%%%%%%%%%%%%%%%%%%%%%%%%%%%%%%%%%%%%%%%%%%%%%%%%%%%%%%%%%%
\begin{abstract}
%%%%%%%%%%%%%%%%%%%%%%%%%%%%%%%%%%%%%%%%%%%%%%%%%%%%%%%%%%%%%%%%%%%
In this paper we investigate asymptotic isotropization. We derive
the asymptotic dynamics of spatially inhomogeneous cosmological
models with a perfect fluid matter source and a positive
cosmological constant near the de Sitter equilibrium state at late
times, and near the flat FL equilibrium state at early times. Our
results show that there exists an open set of solutions approaching
the de Sitter state at late times, consistent with the cosmic
no-hair conjecture.  On the other hand, solutions that approach the
flat FL state at early times are special and admit a so-called
isotropic initial singularity. For both classes of models the
asymptotic expansion of the line element contains an arbitrary
spatial metric at leading order, indicating asymptotic spatial
inhomogeneity. We show, however, that in the asymptotic regimes
this spatial inhomogeneity is significant only at super-horizon
scales.

%%%%%%%%%%%%%%%%%%%%%%%%%%%%%%%%%%%%%%%%%%%%%%%%%%%%%%%%%%%%%%%%%%%
\end{abstract}
%%%%%%%%%%%%%%%%%%%%%%%%%%%%%%%%%%%%%%%%%%%%%%%%%%%%%%%%%%%%%%%%%%%
\begin{flushleft}
PACS number(s): 04.20.-q, 98.80.Jk, 04.20.Dw, 04.20.Ha \\
Preprint number(s): QMUL-AU-2003-002, gr-qc/0306118
\end{flushleft}
%\vfill
%\newpage
%%%%%%%%%%%%%%%%%%%%%%%%%%%%%%%%%%%%%%%%%%%%%%%%%%%%%%%%%%%%%%%%%%%

%%%%%%%%%%%%%%%%%%%%%%%%%%%%%%%%%%%%%%%%%%%%%%%%%%%%%%%%%%%%%%%%%%%
\section{Introduction}
\lb{sec:intro}
%%%%%%%%%%%%%%%%%%%%%%%%%%%%%%%%%%%%%%%%%%%%%%%%%%%%%%%%%%%%%%%%%%%
In a recent paper we presented a general framework for analyzing
the dynamics of generic spatially inhomogeneous cosmological
models, referred to briefly as $G_{0}$ cosmologies (see
Ref.~\ct{uggetal2003}, which we will henceforth refer to as
Paper~I).  We employed Hubble-normalized scale-invariant variables
which were defined within the orthonormal frame formalism, and led
to the formulation of Einstein's field equations with a perfect
fluid matter source as an autonomous system of evolution equations
and constraints. In this paper we show that our framework can be
used to derive asymptotic expansions for spatially inhomogeneous
cosmologies that undergo asymptotic isotropization at late times or
at the initial singularity.

It is part of the folklore of relativistic cosmology that all
ever-expanding cosmological models with a positive cosmological
constant asymptotically approach the de Sitter solution. A
statement to this effect can be found in Gibbons and
Hawking~\ct{gibhaw77}, p.~2739, in connection with a program to
extend to cosmological event horizons certain ideas concerning
thermodynamics and particle creation that had been applied to black
hole event horizons.  This conjecture, the so-called cosmic no-hair
conjecture, was also made by Hawking and Moss~\ct{hawmos82}, p.~36,
after the introduction of the paradigm of cosmic inflation. Shortly
thereafter, Starobinski\v{\i}~\ct{sta83} took the first steps in
describing the asymptotic structure of the metric for spatially
inhomogeneous cosmologies with a perfect fluid matter source and an
effective cosmological constant that approach the de Sitter
solution at late times, showing that the line element contained~8
(the maximum possible number) free functions of the spatial
coordinates. He commented that the asymptotic solution described
``exponentially rapid local isotropization of the universe", and
coined the phrase ``the cosmological constant is the best
`isotropizer'". In the same year, Wald~\ct{wal83} gave a proof of
the conjecture for ever-expanding spatially homogeneous (SH)
cosmologies.  Subsequently Jensen and Stein-Schabes~\ct{jenste87}
extended Wald's proof to spatially inhomogeneous cosmologies,
subject to the restriction that the 3-Ricci curvature scalar is
negative.  The significance of this result is not clear, however,
for the following reason.  The complete set of SH cosmologies can
be divided into one invariant subset for which the 3-Ricci
curvature scalar is never positive (Bianchi Type--I to Type--VIII)
for which Wald's theorem applies, and another invariant subset
containing Bianchi Type--IX and Kantowski--Sachs models, for which
the dynamical behavior of the 3-Ricci curvature scalar is more
complicated (some models approach the de Sitter solution while
others re-collapse). Such a natural division into invariant subsets
does not exist in the $G_{0}$ case. Hence, in $G_{0}$ cosmology the
situation is still unclear as regards possible ``cosmic no-hair
theorems". A variant of such a theorem, however, was proved very
recently by Tchapnda and Rendall~\ct{tchren2003} for cosmologies
with collisionless matter and positive cosmological constant that
are either plane or hyperbolic-plane symmetrical.

The motivation for much of the above work was provided by the idea
of cosmic inflation, in that the de Sitter solution is regarded as
describing the Universe in the late stages of an inflationary
epoch.  On the other hand, the de Sitter solution can also be
regarded as a simple prototype of a cosmological model undergoing
accelerated expansion during the present epoch.  Indeed, recent
observations have focused attention on such models.  In particular,
five years ago two independently working observational groups
reported that the analysis of the type Ia supernova redshift data
they had gathered suggested that in the present epoch the Universe
may be in a state of accelerated expansion (see, e.g., Perlmutter
{\em et al\/}~\ct{peretal98}, Schmidt {\em et al\/}~\ct{schetal98},
and Schmidt~\ct{sch2002}). The acceleration could be due to either
a veritable, repulsive cosmological constant, or some form of
``dark energy''; see Carroll~\ct{car2001} and Ellis~\ct{ell2003}
for further details and references.

In this paper, motivated by the above discussion, we
consider $G_{0}$ cosmologies that are future asymptotic to the de
Sitter solution. We give a formal definition of this notion within
the Hubble-normalized state space and then derive asymptotic
expansions, valid as $t\rightarrow\infty$, for the
Hubble-normalized variables.

The second type of asymptotic isotropization, namely
isotropization at the initial singularity, arises in connection
with the notion of {\em quiescent cosmology\/} (cf., e.g.,
Barrow~\ct{bar78}), which provides an alternative to cosmic
inflation. The idea is that, due to entropy considerations on a
cosmological scale, a suitable initial condition for the Universe
is that the Weyl curvature should be zero (or at least dynamically
unimportant) at the initial singularity; this is the {\em Weyl
curvature hypothesis\/} by Penrose~\ct{pen79}, p.~630. This
hypothesis leads to the notion of an {\em isotropic initial
singularity\/}.  Cosmological initial singularities of this type
first arose as a special case in the general analysis of initial
singularities performed by Lifshitz and Khalatnikov (LK
hereafter)~\ct{lifkha63}, p.~203. This type of initial singularity
was also encountered in the work of Eardley, Liang and
Sachs~\ct{earetal72}, p.~101. Subsequently, Goode and
Wainwright~\ct{goowai85} gave a formal definition of an isotropic
initial singularity, using a conformally related metric, and
derived various properties. Motivated by these ideas, we consider
$G_{0}$ cosmologies that are past asymptotic to the flat FL
solution and give a formal definition of this notion within the
Hubble-normalized state space and then derive asymptotic
expansions, valid as $t\rightarrow -\infty$, for the
Hubble-normalized variables.

In this paper, we also extend the formalism of Paper~I to cover
geodesic null congruences and structures in connection with the
issue of the formation of particle or event horizons (cf.~the
classic paper by Rindler~\ct{rin56}), thus naturally generalizing
the work by Nilsson {\em et al\/}~\ct{niletal2000} from an SH
context to a general spatially inhomogeneous setting.

The plan of this paper is as follows.  In Sec.~\ref{sec:eq}, we
present the Hubble-normalized evolution equations and constraints
for $G_{0}$ cosmologies that arise from Einstein's field equations
and the matter equations.  In Sec.~\ref{sec:ds}, we give the
detailed asymptotic form of the Hubble-normalized variables for
$G_{0}$ cosmologies that approach the de Sitter solution at late
times, and discuss various features of this class of cosmological
models.  In Sec.~\ref{sec:flatfl}, we present analogous results for
$G_{0}$ cosmologies that approach the flat FL solution at early
times.  We conclude in Sec.~\ref{sec:disc} with a discussion of the
analogies between the two classes of cosmological models that
undergo asymptotic isotropization, and raise some issues for future
study. The proofs of the results are given in the appendix.

%%%%%%%%%%%%%%%%%%%%%%%%%%%%%%%%%%%%%%%%%%%%%%%%%%%%%%%%%%%%%%%%%%%
\section{Evolution equations and constraints}
\lb{sec:eq}
%%%%%%%%%%%%%%%%%%%%%%%%%%%%%%%%%%%%%%%%%%%%%%%%%%%%%%%%%%%%%%%%%%%
We consider spatially inhomogeneous cosmological models with a
positive cosmological constant, $\Lambda$, and a perfect fluid
matter source with a linear barotropic equation of state. We thus
have
\be
\lb{eos}
\ti{p}(\ti{\mu}) = (\gam-1)\,\ti{\mu} \ ,
\ee
where $\ti{\mu}$ is the total energy density of the fluid (assumed
to be non-negative) and $\ti{p}$ its isotropic pressure, in the
rest 3-spaces associated with the fluid 4-velocity vector field
$\ti{\vecu}$, while $\gam$ is a constant parameter. The range
\be
\lb{gam}
1 \leq \gam < 2
\ee
is of particular physical interest, since it ensures that the
perfect fluid satisfies the dominant and strong energy conditions
and the causality requirement that the speed of sound should be
less than that of light. The values $\gam = 1$ and
$\gam = \tfrac{4}{3}$ correspond to pressure-free matter
(``dust'') and incoherent radiation, respectively.

We express the fluid 4-velocity vector field by
\be
\ti{\vecu} := \Gamma(\vece_{0}+\vec{v}) \ ,
\ee
where $\vece_{0}$ is a vorticity-free timelike reference
congruence, $\vec{v}$ the peculiar velocity vector field of the
fluid relative to the rest 3-spaces of $\vece_{0}$, and
\be
\lb{jwe5}
\Gamma := \frac{1}{\sqrt{1-v^{2}}} \ , \hspace{10mm}
v^{2} := v_{a} v^{a} \ ,
\ee
defines the usual Lorentz factor. To obtain an orthonormal frame,
$\{\,\vece_{a}\,\}_{a = 0,1,2,3}$, we supplement the vector field
$\vece_{0}$ with an orthonormal spatial frame
$\{\,\vece_{\alpha}\,\}_{\alpha = 1,2,3}$ in the rest 3-spaces of
$\vece_{0}$. The frame metric is then given by $\eta_{ab} =
\mbox{diag}\,[\,-\,1, 1, 1, 1\,]$. To convert the dynamical
equations of the orthonormal frame formalism to a system of partial
differential equations, it is necessary to introduce a set of local
coordinates $\{x^{\mu}\}_{\mu = 0,1,2,3} = \{t,x^{i}\}_{i =
1,2,3}$.

%------------------------------------------------------------------
\subsection{Hubble-normalized Einstein--Euler equations in
  separable volume gauge}
%------------------------------------------------------------------
In Paper~I, we employed the orthonormal frame formalism and
introduced Hubble-normalized scale-invariant frame, connection and
matter variables, as follows.\footnote{We use units such that
Newton's gravitational constant $G$ and the speed of light in
vacuum $c$ are given by $8\pi G/c^{2} = 1$ and $c = 1$. Then frame
and connection variables have physical dimension $\lgth^{-1}$,
while curvature variables have physical dimension $\lgth^{-2}$.}
\bea
\lb{dlfrder}
\{\,\parb_{0}, \,\parb_{\alpha}\,\}
& := & \{\,\vece_{0}, \,\vece_{\alpha}\,\}/H \\
\lb{dlcon}
\{
\,\Udot^{\alpha},
\,A^{\alpha},
\,N_{\alpha\beta},
\,\Sig_{\alpha\beta},
\,R^{\alpha}\,\}
& := & \{
\,\udot^{\alpha},
\,a^{\alpha},
\,n_{\alpha\beta},
\,\sig_{\alpha\beta},
\,\Om^{\alpha}\,\}/H \\
\lb{dlcurv}
\{\,\Om, \,\Oml\,\}
& := & \{\,\mu, \,\Lambda \,\}/(3H^{2}) \ .
\eea
Expressing the Hubble-normalized frame vector fields with respect
to a local coordinate basis leads to: \enl
\be
\lb{dl13cocom}
\parb_{0} = \cn^{-1}\,(\ptl_{t}-N^{i}\,\ptl_{i}) \ , \hspace{10mm}
\parb_{\alpha} = E_{\alpha}{}^{i}\,\ptl_{i} \ ,
\ee
where $\cn$ and $N^{i}$ are viewed as coordinate gauge source
functions, while the $E_{\alpha}{}^{i}$ are dependent variables
that we refer to as the frame variables.

The introduction of Hubble-normalized scale-invariant variables
leaves $H$ as the only dependent variable carrying a physical
dimension, namely $\lgth^{-1}$. The evolution equation for $H$
decouples and is given by
\be
\lb{hq}
\parb_{0}H  =  -\,(q+1)\,H \ ,
\ee
where $q$ is the deceleration parameter, familiar from
observational cosmology.  The expression for $q$, which depends on
the choice of temporal gauge, is given below.  It is also necessary
to introduce the spatial Hubble gradient $r_\alpha$, defined by
\be
\lb{ralpha}
r_{\alpha}  := -\,\frac{1}{H}\,\parb_{\alpha}H \ .
\ee

Throughout this paper, we will employ the {\em separable volume
gauge\/} introduced in Paper~I, which is characterized by the
following choice of coordinate and frame gauge source functions:
\be
\lb{sepvol}
N^{i} = 0 \ ,
\hspace{10mm}
\cn = 1
\hsp5 \Rightarrow \hsp5
\Udot_{\alpha} = r_{\alpha} \ ;
\ee
the choice $\cn = 1$ corresponds to setting the dimensional lapse
function equal to the momentary Hubble radius, $N = H^{-1}$.  With
this choice of temporal gauge the deceleration parameter is given
by
\be
\lb{hdecel}
q = 2\Sig^{2}
+ \tfrac{1}{2}\,G_{+}^{-1}\,[\,(3\gam-2)+(2-\gam)\,v^{2}\,]\,\Om
- \Oml - \tfrac{1}{3}\,(\parb_{\alpha}
-2A_{\alpha})\,r^{\alpha} \ ,
\ee
where $\Sig^{2} = \tfrac16\,(\Sig_{\alpha\beta}\Sig^{\alpha\beta})$,
and the scalars $G_{\pm}$ are defined by (see Eq.~(6) in Paper~I)
\be
\lb{Gpm}
G_{\pm} = 1 \pm (\gam-1)\,v^{2} \ .
\ee
It should be noted that the temporal gauge choice~(\ref{sepvol})
does not determine a unique family of spacelike 3-surfaces ${\cal
S}$:$\{t=\mbox{constant}\}$ and associated normal timelike
reference congruence~$\vece_{0}$.  In Sec.~\ref{sec:flatfl} [\,see
Eq.~(\ref{fl_Hhat})\,], we will use this remaining freedom to
simplify the asymptotic expansions that we derive.  There is also
freedom in the choice of the spatial frame vectors and spatial
coordinates. For the applications in this paper it is desirable to
choose a Fermi-propagated spatial frame, i.e., one which satisfies
\be
\lb{Fermi}
R^\alpha =0 \ .
\ee
The remaining spatial gauge freedom is a time-independent rotation
of the spatial frame vectors,
\be
\lb{rotation}
\tilde{\vece}_{\alpha} = \mathcal{O}_{\alpha}{}^{\beta}(x^{i})
\,\vece_{\beta} \ .
\ee
There is also the following freedom in the choice of spatial
coordinates:
\be
\lb{x_trans}
\tilde{x}^{i} = f^{i}(x^{j}) \ .
\ee

In the separable volume gauge with Fermi-propagated spatial frame, the 
Hubble-normalized state vector for $G_{0}$ cosmologies is given by
\be
\lb{stateX}
\vec{X} = (E_{\alpha}{}^{i}, r_{\alpha}, A^{\alpha},
N_{\alpha\beta}, \Sig_{\alpha\beta}, \Om, v^{\alpha}, \Oml)^{T} \ .
\ee
The evolution equations and constraints that the components of
$\vec{X}$ have to satisfy are given below.\footnote{These equations
are obtained by imposing the restriction (\ref{sepvol}) on
Eqs. (33)--(35), (38), (44), (47), (144) and (145) in Paper~I.  We
keep the $R^\alpha$ in the equations for future reference.} \enl

\noindent
{\em Evolution equations\/}:
\bea
\lb{dl13comts}
\ptl_{t}E_{\alpha}{}^{i}
& = & (q\,\d_{\alpha}{}^{\beta} - \Sig_{\alpha}{}^{\beta}
+ \eps_{\alpha\gam}{}^{\beta}\,R^{\gam})\,E_{\beta}{}^{i} \\
\lb{dlrdot}
\ptl_{t}r_{\alpha}
& = & (q\,\d_{\alpha}{}^{\beta} - \Sig_{\alpha}{}^{\beta}
+ \eps_{\alpha\gam}{}^{\beta}\,R^{\gam})\,r_{\beta}
+ \parb_{\alpha}q \\
\lb{dladot}
\ptl_{t}A^{\alpha}
& = & (q\,\d^{\alpha}{}_{\beta} - \Sig^{\alpha}{}_{\beta}
+ \eps^{\alpha}{}_{\gam\beta}\,R^{\gam})\,A^{\beta}
+ \tfrac{1}{2}\,\parb_{\beta}\,(\Sig^{\alpha\beta}
+\eps^{\alpha\beta}{}_{\gam}\,R^{\gam}) \\
\lb{dlsigdot}
\ptl_{t}\Sig^{\alpha\beta}
& = & (q-2)\,\Sig^{\alpha\beta} - 2N^{\la\alpha}{}_{\gam}\,
N^{\beta\ra\gam} + N_{\gam}{}^{\gam}\,N^{\la\alpha\beta\ra}
- \d^{\gam\la\alpha}\,(\parb_{\gam}-r_{\gam})\,A^{\beta\ra}
\nonumber \\
& & \hsp5 + \ \eps^{\gam\delta\la\alpha}\,[\,(\parb_{\gam}
-2A_{\gam})\,N^{\beta\ra}{}_{\delta}
+ 2R_{\gam}\,\Sig^{\beta\ra}{}_{\delta}\,]
+ (\d^{\gam\la\alpha}\,\parb_{\gam}
+A^{\la\alpha})\,r^{\beta\ra}
+ 3\,\frac{\gam}{G_{+}}\,
\Om v^{\la\alpha}v^{\beta\ra}
\\
\lb{dlndot}
\ptl_{t}N^{\alpha\beta}
& = & (q\,\d^{(\alpha}{}_{\delta}
+ 2\Sig^{(\alpha}{}_{\delta}
+ 2\eps^{\gam}{}_{\delta}{}^{(\alpha}\,R_{\gam})\,N^{\beta)\delta}
- \parb_{\gam}\,
(\eps^{\gam\delta(\alpha}\,\Sig^{\beta)}{}_{\delta}
- \d^{\gam(\alpha}\,R^{\beta)} + \d^{\alpha\beta}\,R^{\gam}) \\
\lb{dlomdot}
\ptl_{t}\Om
& = & -\,\frac{\gam}{G_{+}}\,v^{\alpha}\,\parb_{\alpha}\Om
+ G_{+}^{-1}\,[\,2G_{+}q - (3\gam-2) - (2-\gam)\,v^{2}
- \gam\,(\Sig_{\alpha\beta}v^{\alpha}v^{\beta}) \nonumber \\
& & \hspace{55mm} - \ \gam\,(\parb_{\alpha}-2A_{\alpha})\,v^{\alpha}
+ \gam\,v^{\alpha}\,\parb_{\alpha}\ln G_{+}\,]\,\Om \\
\lb{dlvdotf}
\ptl_{t}v^{\alpha}
& = & -\,v^{\beta}\,\parb_{\beta}v^{\alpha}
+ \d^{\alpha\beta}\,\parb_{\beta}\ln G_{+}
- \frac{(\gam-1)}{\gam}\,(1-v^{2})\,\d^{\alpha\beta}\,
(\parb_{\beta}\ln\Om-2r_{\beta}) \nonumber \\
& & + \ G_{-}^{-1}\,\Big[\,(\gam-1)\,(1-v^{2})\,(\parb_{\beta}
v^{\beta}) - (2-\gam)\,v^{\beta}\,\parb_{\beta}\ln G_{+}
\nonumber \\
& & \hspace{15mm}
+ \ \frac{(\gam-1)}{\gam}\,(2-\gam)\,(1-v^{2})\,v^{\beta}\,
(\parb_{\beta}\ln\Om-2r_{\beta})
+ (3\gam-4)\,(1-v^{2}) \nonumber \\
& & \hspace{15mm} + \ (2-\gam)\,(\Sig_{\beta\gam}v^{\beta}
v^{\gam}) + G_{-}\,(r_{\beta}v^{\beta})
+ [G_{+}-2(\gam-1)]\,(A_{\beta}v^{\beta})\,\Big]\,v^{\alpha}
\nonumber \\
& & - \ \Sig^{\alpha}{}_{\beta}\,v^{\beta}
+ \eps^{\alpha}{}_{\beta\gam}\,R^{\beta}\,v^{\gam}
- r^{\alpha} - v^{2}\,A^{\alpha}
+ \eps^{\alpha\beta\gam}\,N_{\beta\delta}\,v_{\gam}\,v^{\delta}
\\
\lb{dlomldot}
\ptl_{t}\Oml
& = & 2\,(q+1)\,\Oml \ .
\eea
where $q$, $G_{\pm}$ and $\parb_{\alpha}$ are given by
Eqs.~(\ref{hdecel}), (\ref{Gpm}) and (\ref{dl13cocom}).  \enl

\noindent
{\em Constraints\/}:
\bea
\lb{dl13comss}
0 & = & ({\cal C}_{{\rm com}})^{i}{}_{\alpha\beta}
:= 2\,(\parb_{[\alpha}-r_{[\alpha}-A_{[\alpha})\,
E_{\beta]}{}^{i}
- \eps_{\alpha\beta\delta}\,N^{\delta\gam}\,E_{\gam}{}^{i} \\
\lb{dlgauss}
0 & = & ({\cal C}_{\rm G})
\ := \ 1 - \ck - \Sig^{2} - \Om - \Oml \\
\lb{dlcodacci}
0 & = & ({\cal C}_{\rm C})^{\alpha}
\ := \ \parb_{\beta}\Sig^{\alpha\beta}
+ (2\d^{\alpha}{}_{\beta}-\Sig^{\alpha}{}_{\beta})\,r^{\beta}
- 3A_{\beta}\,\Sig^{\alpha\beta}
- \eps^{\alpha\beta\gam}\,N_{\beta\delta}\,\Sig_{\gam}{}^{\delta}
+ 3\,\frac{\gam}{G_{+}}\,\Om v^{\alpha} \\
\lb{dljacobi1}
0 & = & ({\cal C}_{\rm J})^{\alpha}
\ := \ (\parb_{\beta}-r_{\beta})\,(N^{\alpha\beta}
+\eps^{\alpha\beta\gam}\,A_{\gam}) - 2A_{\beta}\,N^{\alpha\beta}
\\
\lb{dloml}
0 & = & ({\cal C}_{\Lambda})_{\alpha}
\ := \ (\parb_{\alpha} - 2r_{\alpha})\,\Oml \ .
\eea

\noindent
{\em Auxiliary 3-curvature variables\/}:
\bea
\lb{omkdef}
\ck & := & -\,\tfrac{1}{3}\,(2\parb_{\alpha}
-2r_{\alpha}-3A_{\alpha})\,A^{\alpha}
+ \tfrac{1}{6}\,(N_{\alpha\beta}N^{\alpha\beta})
- \tfrac{1}{12}\,(N_{\alpha}{}^{\alpha})^{2} \\
\lb{dltrfr3ric}
{\cal S}_{\alpha\beta}
& := & -\,\tfrac{1}{3}\,\eps^{\gam\delta}{}_{\la\alpha}\,
(\parb_{|\gam|}-r_{|\gam|}-2A_{|\gam|})\,N_{\beta\ra\delta}
+ \tfrac{1}{3}\,(\parb_{\la\alpha}-r_{\la\alpha})\,A_{\beta\ra}
+ \tfrac{2}{3}\,N_{\la\alpha}{}^{\gam}\,N_{\beta\ra\gam}
- \tfrac{1}{3}\,N_{\gam}{}^{\gam}\,N_{\la\alpha\beta\ra} \\
\lb{dl3ct}
{\cal C}_{\alpha\beta} & := & \eps^{\gam\delta}{}_{\la\alpha}\,
(\parb_{|\gam|}-2r_{|\gam|}-A_{|\gam|})\,{\cal S}_{\beta\ra\delta}
- 3N_{\la\alpha}{}^{\gam}\,{\cal S}_{\beta\ra\gam}
+ \tfrac{1}{2}\,N_{\gam}{}^{\gam}\,{\cal S}_{\alpha\beta} \ ,
\eea
where ${\cal S}_{\alpha\beta}$ and $\ck$ are the tracefree part and
trace part of the Hubble-normalized 3-Ricci curvature of a
spacelike 3-surface ${\cal S}$:$\{t=\mbox{constant}\}$,
respectively, while ${\cal C}_{\alpha\beta}$ is the
Hubble-normalized 3-Cotton--York tensor which provides information
on the conformal curvature properties of a spacelike 3-surface
${\cal S}$:$\{t=\mbox{constant}\}$.

The 3-Ricci curvature variables ${\cal S}_{\alpha\beta}$ and $\ck$
satisfy the Hubble-normalized twice-contracted 3-Bianchi identity,
given by
\be
\lb{dl3bianid}
0 \equiv (\parb_{\beta}-2r_{\beta}-3A_{\beta})\,{\cal S}^{\alpha\beta}
- \eps^{\alpha\beta\gam}\,N_{\beta\delta}\,{\cal S}_{\gam}{}^{\delta}
+ \tfrac{1}{3}\,\d^{\alpha\beta}\,(\parb_{\beta}-2r_{\beta})\,\ck \ .
\ee
%

%------------------------------------------------------------------
\subsection{Energy-normalized null geodesics augmentation}
%------------------------------------------------------------------
A few years ago, Nilsson {\em et al\/}~\ct{niletal2000} extended
the Hubble-normalized equations for SH cosmology by adding the
equations for an energy-normalized geodesic (timelike or null)
congruence. This strategy turned out to be very useful, both from
an analytical and a numerical point of view. We therefore
generalize that approach to the present $G_{0}$ cosmology setting,
and this will subsequently allow us to investigate the connection
between asymptotic silence of the gravitational field dynamics in
relativistic cosmology and the formation of particle or event
horizons (see Paper I, Sec.~4.1).

Instead of being interested in the null geodesics emanating from a
single given event, we will be interested in all possible null
geodesics. We will thus not consider the problem of integrating a
system of ordinary differential equations (associated with the
former situation), but will instead consider all possible null
geodesic vector fields (which, at a later stage, can be integrated
to yield all possible associated flows). We therefore start by
considering a vector field $k^{a}$ which is tangent to a geodesic
null congruence. This vector field is thus governed by the
equations
\bea
\lb{geodeq}
k^{b}\nabla_{b}k^{a} & = & 0 \\
\lb{nullcon}
k_{a}k^{a} & = & 0 \ .
\eea
We now make a (3+1)-split of Eq.~(\ref{geodeq}) with respect to a
general orthonormal frame. Using the notation of Paper~I and
Ref.~\ct{niletal2000}, this yields
\bea
\lb{edot}
{\cal E}\,\vece_{0}({\cal E})
& = & -\,k^{\alpha}\,(\vece_{\alpha}+\udot_{\alpha})\,({\cal E})
- (H\,\d_{\alpha\beta}+\sig_{\alpha\beta})\,k^{\alpha}k^{\beta} \\
\lb{kadot}
{\cal E}\,\vece_{0}(k^{\alpha})
& = & -\,k^{\beta}\,\vece_{\beta}(k^{\alpha})
- {\cal E}\,(H\,\d^{\alpha}{}_{\beta}+\sig^{\alpha}{}_{\beta})\,
k^{\beta} - (k_{\beta}k^{\beta})\,a^{\alpha}
+ (k^{\beta}a_{\beta})\,k^{\alpha}
- \eps^{\alpha}{}_{\beta\delta}\,n_{\gam}{}^{\delta}\,
k^{\beta}k^{\gam} \nonumber \\
& & \hsp5 - \ {\cal E}^{2}\,\udot^{\alpha}
+ {\cal E}\,\eps^{\alpha}{}_{\beta\gam}\,\Om^{\beta}\,k^{\gam} \ ,
\eea
while a (3+1)-split of Eq.~(\ref{nullcon}) yields
\be
\lb{nullcon2}
-\,{\cal E}^{2} + (k_{\alpha}k^{\alpha}) = 0 \ .
\ee
Here ${\cal E} := k^{0}$ denotes the energy (frequency) of the
particles that are traveling along the geodesic null congruence.

We now employ the Hubble-normalized gravitational field variables
given in Eqs.~(\ref{dlfrder}) and~(\ref{dlcon}), and in addition
introduce {\em energy-normalized\/} null vector variables as in
Ref.~\ct{niletal2000} according to
\be
\lb{dlka}
K^{\alpha} := \frac{k^{\alpha}}{{\cal E}} \ .
\ee
With Eq.~(\ref{nullcon2}) we thus find
\be
\lb{dlkamag}
K_{\alpha}K^{\alpha} = 1 \ ,
\ee
where the $K^{\alpha}$ correspond to the direction cosines of the
null geodesics.

Energy-normalization implies that the dimensional
equation~(\ref{edot}) for ${\cal E}$ decouples and can be written
as
\be
\lb{sdef}
\parb_{0}{\cal E} = -\,(s+1)\,{\cal E} \ ,
\ee
where $s$ is given by
\be
\lb{sexpr}
s := (\Sig_{\alpha\beta}K^{\alpha}K^{\beta})
- (t_{\alpha}-\Udot_{\alpha})\,K^{\alpha} \ ,
\ee
which is obtained from Eqs.~(\ref{edot}), (\ref{dlka}),
(\ref{dlkamag}) and~(\ref{dlcon}).  It is also necessary to
introduce the {\em spatial energy gradient\/} $t_\alpha$, defined
by
\be
\lb{tadef}
t_{\alpha} := -\,\frac{1}{\cal E}\,\parb_{\alpha}{\cal E} \ ,
\ee
and governed by
\bea
\lb{dltadot}
\parb_{0}t_{\alpha}
& = & (q\,\d_{\alpha}{}^{\beta} - \Sig_{\alpha}{}^{\beta}
+ \eps_{\alpha\gam}{}^{\beta}\,R^{\gam})\,t_{\beta}
+ (\parb_{\alpha}-r_{\alpha}+\Udot_{\alpha})\,(s+1) \\
\lb{dltacon}
0 & = & ({\cal C}_{t})^{\alpha}
\ := \ [\,\eps^{\alpha\beta\gam}\,
(\parb_{\beta}-r_{\beta}-A_{\beta})
- N^{\alpha\gam}\,]\,t_{\gam} \ .
\eea
The latter equations are obtained by choosing $f = {\cal E}$ in the
commutator equations~(31) and (32) of Paper~I, and making use of
Eqs.~(\ref{sdef}) and (\ref{tadef}) given
above. Equations~(\ref{dltadot}) and~(\ref{dltacon}) constitute
integrability conditions for Eqs.~(\ref{sdef}) and (\ref{tadef}).

We now write the evolution equation~(\ref{kadot}) in
energy/Hubble-normalized form.
\bea
\lb{dlkadot}
\parb_{0}K^{\alpha}
& = & -\,K^{\beta}\,(\parb_{\beta}-t_{\beta}-A_{\beta})\,K^{\alpha}
+ (s\,\d^{\alpha}{}_{\beta}-\Sig^{\alpha}{}_{\beta})\,K^{\beta}
- A^{\alpha} - \eps^{\alpha}{}_{\beta\delta}\,N_{\gam}{}^{\delta}\,
K^{\beta}K^{\gam} \nonumber \\
& & \hsp5 - \ \Udot^{\alpha}
+ \eps^{\alpha}{}_{\beta\gam}\,R^{\beta}\,K^{\gam} \ .
\eea
Note that the contraction of Eq.~(\ref{dlkadot}) with $K_{\alpha}$
vanishes identically on account of Eqs.~(\ref{dlkamag})
and~(\ref{sexpr}).

Let us now consider these equations in the separable volume gauge,
specified by Eqs.~(\ref{sepvol}). Then
\bea
\lb{Keq}
\ptl_{t}K^{\alpha}
& = & -\,K^{\beta}\,(\parb_{\beta}
-t_{\beta}-A_{\beta})\,K^{\alpha}
+ (s\,\d^{\alpha}{}_{\beta}-\Sig^{\alpha}{}_{\beta})\,K^{\beta}
- A^{\alpha} - \eps^{\alpha}{}_{\beta\delta}\,N_{\gam}{}^{\delta}\,
K^{\beta}K^{\gam} \nonumber \\
& & \hsp5 - \ r^{\alpha}
+ \eps^{\alpha}{}_{\beta\gam}\,R^{\beta}\,K^{\gam} \\
\lb{teq}
\ptl_{t}t_{\alpha}
& = & (q\,\d_{\alpha}{}^{\beta} - \Sig_{\alpha}{}^{\beta}
+ \eps_{\alpha\gam}{}^{\beta}\,R^{\gam})\,t_{\beta}
+ \parb_{\alpha}s \ ,
\eea
where
\be
\lb{s}
s = (\Sig_{\alpha\beta}K^{\alpha}K^{\beta})
- (t_{\alpha}-r_{\alpha})\,K^{\alpha} \ .
\ee
Once the energy-normalized null vector variables $K^{\alpha}$ have
been obtained, one can solve for the associated geodesic null
congruence by integrating the relation
\be
\lb{geo}
\frac{{\rm d}x^{i}}{{\rm d}t} = E_{\alpha}{}^{i}\,K^{\alpha} \ ,
\ee
subject to appropriate initial conditions.

In this paper, we will use the above equations to determine
asymptotic geodesic null structures and derive properties
concerning particle and event horizons which are associated with
asymptotic silence.

%%%%%%%%%%%%%%%%%%%%%%%%%%%%%%%%%%%%%%%%%%%%%%%%%%%%%%%%%%%%%%%%%%%
\section{de Sitter-like future asymptotics}
\lb{sec:ds}
%%%%%%%%%%%%%%%%%%%%%%%%%%%%%%%%%%%%%%%%%%%%%%%%%%%%%%%%%%%%%%%%%%%
In this section we give the detailed asymptotic form of the
Hubble-normalized variables for $G_{0}$ cosmologies that approach
the de Sitter solution at late times.

%------------------------------------------------------------------
\subsection{$G_{0}$ cosmologies future asymptotic to the de Sitter
  solution}
\lb{subsec:ds}
%------------------------------------------------------------------
The de Sitter solution of Einstein's field equations describes a
cosmological model with zero matter energy density and positive
cosmological constant, which is undergoing exponential expansion.
Employing a (3+1)-decomposition with intrinsically flat spacelike
3-surfaces, the line element and the associated timelike reference
congruence are given by
\begin{gather}
\lb{ds_metric}
{\rm d}s^{2} = -\,{\rm d}T^{2}
+ \ell_{0}^{2}\,\e^{2\sqrt{\Lambda/3}\,T}\,
({\rm d}x^{2} + {\rm d}y^{2} + {\rm d}z^{2}) \ ,
\\
\vece_{0} = \ptl_{T} \ ,
\end{gather}
where $\Lambda > 0$ is the cosmological constant, and $T$ is clock
time. It is convenient to choose the unit of $\lgth$ as
follows:
\be
\lb{ds_l0}
\ell_{0} = \sqrt{\frac{3}{\Lambda}} \ .
\ee
The Hubble scalar is constant and positive, and is given by
\be
\lb{ds_H}
H = \sqrt{\frac{\Lambda}{3}} = \ell_{0}^{-1} \ .
\ee
It follows from Eq.~(\ref{hq}) that the deceleration parameter is
constant and negative, being given by
\be
q = -\,1 \ .
\ee
We also find it convenient to use a conformal time coordinate
$\Eta$, defined by
\be
\lb{ds_eta}
\Eta := \e^{-T/\ell_{0}} \ .
\ee
The line element is then cast into the form
\be
{\rm d}s^{2} = \ell_{0}^{2}\,\Eta^{-2}\,(-\,{\rm d}\Eta^{2}
+ {\rm d}x^{2} + {\rm d}y^{2} + {\rm d}z^{2}) \ .
\ee

Relative to the natural orthonormal frame associated with the line
element~(\ref{ds_metric}), the Hubble-normalized variables are
\begin{gather}
\lb{ds_1}
E_{\alpha}{}^{i} = \e^{-t}\,\delta_{\alpha}{}^{i}\ ,
\hspace{10mm}
r_{\alpha} = 0 \ ,
\\
\Sig_{\alpha\beta} = 0 \ , \hspace{10mm}
A^{\alpha} = 0 \ , \hspace{10mm}
N_{\alpha\beta} = 0 \ ,
\\
\lb{ds_3}
\Om = 0 \ , \hspace{10mm} \Oml = 1 \ ,
\end{gather}
where the dynamical time coordinate $t$ is introduced via
\be
t = \frac{T}{\ell_{0}} \ ,
\ee
or, equivalently,
\be
\Eta = \e^{-t} \ .
\ee
We note that the volume density $\cv$ is given by
\be
\cv = \ell_{0}^{3}\,\e^{3t} \ ,
\ee
so that the separable volume gauge conditions are satisfied.

Observe that the frame variables satisfy
$\lim_{t\rightarrow\infty}E_{\alpha}{}^{i} = 0$.  Thus, the de
Sitter solution is described by an orbit in the Hubble-normalized
state space that is future asymptotic to an equilibrium point on
the silent boundary, $E_{\alpha}{}^{i} = 0$.

Motivated by Eqs.~(\ref{ds_1})--(\ref{ds_3}), we say that {\em a
$G_{0}$ cosmology is future asymptotic to the de Sitter solution\/}
if the following limits are satisfied:
\begin{gather}
\lb{ds_c1}
\lim\limits_{t\rightarrow\infty}(E_{\alpha}{}^{i},
r_{\alpha})^{T} = \vec{0} \\
\lim\limits_{t\rightarrow\infty}(\Sig_{\alpha\beta},
A^{\alpha}, N_{\alpha\beta})^{T} = \vec{0} \\
\lb{ds_c3}
\lim\limits_{t\rightarrow\infty} \Om = 0 \ , \hspace{10mm}
\lim\limits_{t\rightarrow\infty} \Oml = 1 \ .
\end{gather}
We note that for the de Sitter solution the peculiar velocity
variables $v^{\alpha}$ are unrestricted, since the Codacci
constraint~(\ref{dlcodacci}) (which may be viewed as determining
the $v^{\alpha}$ algebraically provided that $\Om\neq0$) yields no
information. We will show that if the limits
(\ref{ds_c1})--(\ref{ds_c3}) are satisfied, together with certain
technical restrictions, then the evolution equations determine the
asymptotic form of the $v^{\alpha}$ as $t \rightarrow \infty$.
Wald's theorem~\ct{wal83}, mentioned in the introduction, asserts
that any SH cosmology with a positive cosmological constant,
except those of Kantowski-Sachs and Bianchi Type--IX which may
re-collapse, is future asymptotic to the de Sitter solution, in
the above sense. As discussed in the introduction, at present it
is not known whether Wald's theorem can be generalized to $G_{0}$
cosmologies. We now consider the class of $G_{0}$ cosmologies
which are future asymptotic to the de Sitter solution, and present
the asymptotic form of the Hubble-normalized variables as $t
\rightarrow \infty$.

%------------------------------------------------------------------
\subsection{Asymptotic expansions}
\lb{subsec:ds_ae}
%------------------------------------------------------------------
In App.~\ref{app:ds}, we prove that the de Sitter solution is
asymptotically stable, and derive the asymptotic form of the
Hubble-normalized variables for $G_{0}$ cosmologies that are future
asymptotic to the de Sitter solution, in the sense of
Eqs.~(\ref{ds_c1})--(\ref{ds_c3}). It is necessary to impose
certain restrictions on the spatial derivatives; these are given in
App.~\ref{app:ds}.

We now give the asymptotic expansions, using the convention that
``hatted" coefficients depend only on the spatial coordinates:
\bea
\lb{dS_bigO}
\left(E_{\alpha}{}^{i}, A^{\alpha}, N^{\alpha\beta}\right)^{T}
& = &
\left(\hat{E}_{\alpha}{}^{i},
\hat{A}^{\alpha}, \hat{N}^{\alpha\beta}\right)^{T}\e^{-t}
+ {\cal O}(\e^{-3t}) \\
\lb{dSdlr}
r_{\alpha}
& = & -\,\tfrac{1}{2}\,(\hat{E}_{\alpha}{}^{i}\,
\ptl_{i}\hat{\Om}_{k})\,\e^{-3t}
+ {\cal O}(\e^{-(1+3\gam)t}+\e^{-5t}) \\
\lb{dSdlSig}
\Sig^{\alpha\beta}
& = & -\,3\hat{\cal S}^{\alpha\beta}\,\e^{-2t}
+ \hat{\Sig}^{\alpha\beta}\,\e^{-3t} + {\cal O}(\e^{-4t}) \\
\lb{dSdloml}
\Oml
& = & 1 - \hat{\Om}_{k}\,\e^{-2t} + {\cal O}(\e^{-3\gam t}
+ \e^{-4t}) \\
\lb{dSdlom}
\Om & = &
\begin{cases}
\hat{\Om}\,\e^{-3\gam t} + {\cal O}(\e^{(3\gam-8)t})
& \hsp5 \text{for $1 \leq \gam < \tfrac{4}{3}$} \\
\hat{\Om}\,\e^{-4t} + {\cal O}(\e^{-5t})
& \hsp5 \text{for $\gam=\tfrac{4}{3}$} \\
\hat{\Om}\,\e^{-4t} + {\cal O}(\e^{-5t}
+\e^{(-4-2\frac{(3\gam-4)}{(2-\gam)})t})
& \hsp5 \text{for $\tfrac{4}{3} < \gam < 2$}
\end{cases} \\
v^{\alpha} & = &
\begin{cases}
\hat{v}^{\alpha}\,\e^{-t} + {\cal O}(\e^{-3t})
& \hsp5 \text{for $\gam = 1$} \\
\hat{v}^{\alpha}\,\e^{(3\gam-4)t} + {\cal O}(\e^{-t}
+\e^{3(3\gam-4)t})
& \hsp5 \text{for $1 < \gam < \tfrac{4}{3}$} \\
\hat{v}^{\alpha} + {\cal O}(\e^{-t})
& \hsp5 \text{for $\gam = \tfrac{4}{3}$}
\end{cases} \\
\lb{dS_bigO_end}
1-v^{2} & = & \exp\left[\,-\,2\,\frac{(3\gam-4)}{(2-\gam)}\,t\,\right]
\left[\,(1-\hat{v}^{2})
+ {\cal O}(\e^{-t}+\e^{-2\frac{(3\gam-4)}{(2-\gam)}t})\,\right]
\hsp5 \text{for $\tfrac{4}{3} < \gam < 2$} \ .
\eea
The coefficients $\hat{A}^{\alpha}$ and $\hat{N}^{\alpha\beta}$ are
determined by $\hat{E}_{\alpha}{}^{i}$ according to
\begin{align}
\lb{Ahat}
\hat{A}_{\alpha} &= \hat{E}^{\beta}{}_{i}\,
\hat{E}_{[\alpha}{}^{j}\,\ptl_{j}\hat{E}_{\beta]}{}^{i} \\
\label{Nhat}
\hat{N}^{\alpha\beta} &= \hat{E}^{(\alpha}{}_{i}\,
\eps^{\beta)\gam\delta}\,\hat{E}_{\gam}{}^{j}\,
\ptl_{j}\hat{E}_{\delta}{}^{i} \ ,
\end{align}
where the matrix $\hat{E}^{\alpha}{}_{i}$ is the inverse of
$\hat{E}_{\alpha}{}^{i}$:
\be
\lb{hateinv}
\hat{E}^{\alpha}{}_{i}\,\hat{E}_{\beta}{}^{i}
= \d^{\alpha}{}_{\beta} \ , \hspace{10mm}
\hat{E}^{\alpha}{}_{i}\,\hat{E}_{\alpha}{}^{j}
= \d_{i}{}^{j} \ .
\ee
The coefficients $\hat{\Om}_{k}$ and $\hat{\cal S}_{\alpha\beta}$
in Eqs.~(\ref{dSdlr})--(\ref{dSdloml}) are the leading order
coefficients in the asymptotic expansions of the 3-Ricci curvature
variables,
\bea
\Om_{k} & = & \e^{-2t}\left[\,\hat{\Om}_{k}
+ \bigO(\e^{-2t})\,\right] \\
{\cal S}_{\alpha\beta} & = & \e^{-2t}
\left[\,\hat{\cal S}_{\alpha\beta} + \bigO(\e^{-2t})\,\right] \ .
\eea
They are determined by $\hat{E}_{\alpha}{}^{i}$, $\hat{A}^{\alpha}$
and $\hat{N}_{\alpha\beta}$ according to:\footnote{These equations
are the ``hatted" versions of Eqs.~(\ref{omkdef})
and~(\ref{dltrfr3ric}), with $r_{\alpha}$ set to zero.}
\bea
\lb{Omkhat}
\hat{\Om}_{k}
& = & -\,\tfrac{1}{3}\,(2\hat{E}_{\alpha}{}^{i}\ptl_i
-3\hat{A}_{\alpha})\,\hat{A}^{\alpha} +
\tfrac{1}{6}\,(\hat{N}_{\alpha\beta}\hat{N}^{\alpha\beta})
- \tfrac{1}{12}\,(\hat{N}_{\alpha}{}^{\alpha})^{2} \\
\lb{Shat}
\hat{\mathcal{S}}_{\alpha\beta}
& = & -\,\tfrac{1}{3}\,\eps^{\gam\delta}{}_{\la\alpha}\,
(\hat{E}_{|\gam|}{}^{i}\ptl_i-2\hat{A}_{|\gam|})\,\hat{N}_{\beta\ra\delta}
+ \tfrac{1}{3}\,\hat{E}_{\la\alpha}{}^{i}\ptl_i\,\hat{A}_{\beta\ra}
+ \tfrac{2}{3}\,\hat{N}_{\la\alpha}{}^{\gam}\,\hat{N}_{\beta\ra\gam}
- \tfrac{1}{3}\,\hat{N}_{\gam}{}^{\gam}\,\hat{N}_{\la\alpha\beta\ra} \ .
\eea
The constraint $({\cal C}_{\rm C})^{\alpha}$ at order $\e^{-4t}$
provides the restriction for $\hat{v}^{\alpha}$:
\be
\lb{dS_CC}
0 = (\hat{E}_{\beta}{}^{i}\,\ptl_{i}-3\hat{A}_\beta)\,
\hat{\Sig}^{\alpha\beta} - \eps^{\alpha\beta\gam}\,
\hat{N}_{\beta\d}\,\hat{\Sig}_{\gam}{}^{\d}
+ C^{\alpha} + \hat{Q}^{\alpha} \ ,
\ee
where
\be
\lb{dS_CC2}
C^{\alpha} = \begin{cases}
    -\,\d^{\alpha\beta}\,\hat{E}_{\beta}{}^{i}\,\ptl_{i}\hat{\Om}
    & \hsp5 \text{for $\gam = 1$,} \\
    0
    & \hsp5 \text{for $1< \gam < 2$,}
\end{cases} \hspace{10mm}
\hat{Q}^{\alpha} =
\begin{cases}
3\gam\hat{\Om}\hat{v}^{\alpha}
& \hsp5 \text{for $1 \leq \gam < \tfrac{4}{3}$.} \\
12\hat{\Om}\hat{v}^{\alpha}/(3+\hat{v}_{\beta}\hat{v}^{\beta})
& \hsp5 \text{for $\gam = \tfrac{4}{3}$.} \\
3\hat{\Om}\hat{v}^{\alpha}
& \hsp5 \text{for $\tfrac{4}{3} < \gam < 2$.}
\end{cases}
\ee
The constraints $({\cal C}_{\Lambda})_{\alpha}$ and $({\cal C}_{\rm
G})$ have been used in App.~\ref{app:ds} to give the coefficients
in Eqs.~(\ref{dSdlr}) and (\ref{dSdloml}),
respectively.\footnote{Note that up to (and including) order
$\e^{-2t}$, the present expansion also satisfies the gauge
conditions defining the constant mean curvature ($r_{\alpha} = 0$)
and synchronous ($\Udot_{\alpha} = 0$) temporal gauges,
respectively.}

%------------------------------------------------------------------
\subsection{Features of asymptotic to de Sitter $G_{0}$
  cosmologies}
\lb{subsec:ds_prop}
%------------------------------------------------------------------
In this section we investigate certain features of asymptotic to de
Sitter $G_{0}$ cosmologies, as described by the asymptotic
expansions (\ref{dS_bigO})--(\ref{dS_bigO_end}). Our first goal is
to show that these expansions represent a general class of perfect
fluid $G_{0}$ cosmologies. We accomplish this by showing that the
expansions (\ref{dS_bigO})--(\ref{dS_bigO_end}) contain 8 freely
specifiable functions of the spatial coordinates $x^{i}$, the same
number that appear in the initial data for a general cosmological
solution of Einstein's field equations with a perfect fluid matter
source (see LK, p.~188).

%---------------------------------------------
\subsubsection*{Essential arbitrary functions}
%---------------------------------------------
The nine coefficients $\hat{E}_{\alpha}{}^{i}$ in the expansions
(\ref{dS_bigO})--(\ref{dS_bigO_end}) can be chosen as suitably
differentiable arbitrary functions of the~$x^{i}$, which then
successively determine $\hat{A}{}^{\alpha}$,
$\hat{N}_{\alpha\beta}$, $\hat{\Omega}_k$ and
$\hat{S}_{\alpha\beta}$ via Eqs.~(\ref{Ahat}), (\ref{Nhat}),
(\ref{Omkhat}) and (\ref{Shat}).  The five shear rate coefficients
$\hat{\Sig}_{\alpha\beta}$ and the coefficient $\hat{\Om}$ in the
density parameter can also be chosen arbitrarily, while the
peculiar velocity coefficients~$\hat{v}^{\alpha}$ are determined
algebraically by the Codacci constraint (\ref{dS_CC}), employing
Eq.~(\ref{dS_CC2}).  As mentioned in Sec.~\ref{sec:eq}, the
separable volume gauge (\ref{sepvol}) leaves a freedom in the
choice of the 1-parameter family of spacelike 3-surfaces ${\cal
S}$:$\{t=\mbox{constant}\}$.  A coordinate transformation of the
form\footnote{See Starobinski\v{\i}~\ct{sta83} for the analogous
synchronous gauge-preserving transformation.  The function
$\varphi(x^{i})$ can be interpreted in terms of a boost of the
original frame by the boost function $w^{\alpha}(t,x^{i})$ via
$w^{\alpha} =
-\,\d^{\alpha\beta}\,(\hat{E}_{\beta}{}^{i}\,\ptl_{i}\varphi)
\,\e^{-t}+ \bigO(\e^{-2t})$.}
\be
\tilde{t} = t + \varphi(x^{i}) + \bigO(\e^{-2t})\ , \hspace{10mm}
\tilde{x}{}^{i} = x^{i} - \tfrac{1}{2}\left(
\frac{\Lambda}{3}\right)^{-1} \hat{g}^{ij} \ptl_j \varphi\, \e^{-2t}
+ \bigO(\e^{-3t})\ ,
\ee
where $\hat{g}{}^{ij}= \ell_{0}^{-2}\,\d^{\alpha\beta}\,{\hat
E}_{\alpha}{}^{i}\,{\hat E}_{\beta}{}^{j}$, preserves the separable
volume gauge to leading order.  This freedom can be used to fix $\hat{\Om}$,
although we do not choose to do so in general.  Three of the
$\hat{E}_{\alpha}{}^{i}$ can be eliminated by the frame
rotations~(\ref{rotation}), and three by the spatial coordinate
transformations~(\ref{x_trans}), leaving three essentially
arbitrary functions in the $\hat{E}_{\alpha}{}^{i}$. The asymptotic
expansions~(\ref{dS_bigO})--(\ref{dS_bigO_end}) thus contain~8
essentially arbitrary functions of the spatial coordinates. That
is, they represent a general class of perfect fluid $G_{0}$
cosmologies that are future asymptotic to the de Sitter solution.

%--------------------------------
\subsubsection*{Metric expansion}
%--------------------------------
We now derive an asymptotic expansion for the spacetime metric in
order to relate our results to the work of other researchers, in
particular that of Starobinski\v{\i}~\ct{sta83}. We find that the
metric expansion sheds light on the nature of large-scale spatial
inhomogeneity in $G_{0}$ cosmologies that are future asymptotic to
the de Sitter solution.

In the separable volume gauge, the line element has the form
\be
\lb{sepvol_metric}
{\rm d}s^{2} = -\,H^{-2}\,{\rm d}t^{2}
+ g_{ij}\,{\rm d}x^{i}\,{\rm d}x^{j} \ ,
\ee
where [\,see Paper I, Eq. (163)\,]
\be
\lb{g_def}
g_{ij} = H^{-2}\,\d_{\alpha\beta}\,E^{\alpha}{}_{i}\,
E^{\beta}{}_{j} \ .
\ee
Equations~(\ref{dS_bigO})--(\ref{dS_bigO_end}) are substituted into
Eq.~(\ref{hdecel}) to give an improved expansion for $q$, and then
into Eq.~(\ref{dl13comts}) to give an expansion for
$\ptl_{t}E_{\alpha}{}^{i}$, which can be integrated. In the case
$\gam < \tfrac{4}{3}$, we obtain
\be
\lb{E_ae}
E_{\alpha}{}^{i} = \hat{E}_{\beta}{}^{i}\,\e^{-t}
\left[\,\d_{\alpha}{}^{\beta} - \tfrac{1}{2}\left(
3\hat{\mathcal{S}}_{\alpha}{}^{\beta}
+\hat{\Om}_{k}\,\d_{\alpha}{}^{\beta}
\right)\e^{-2t}
+ \tfrac{1}{3}\,\hat{\Sig}_{\alpha}{}^{\beta}\,\e^{-3t}
- \tfrac{1}{2}\,\hat{\Om}\,\d_{\alpha}{}^{\beta}\,
\e^{-3\gam t} + \bigO(\e^{-4t})\,\right] \ .
\ee
The components $E^{\alpha}{}_{i}$ of the corresponding
Hubble-normalized 1-forms form the inverse of the matrix
$E_{\alpha}{}^{i}$, and are given by
\be
\lb{Ein_ae}
E^{\alpha}{}_{i} = \hat{E}^{\beta}{}_{i}\,\e^{t}\left[\,
\d^{\alpha}{}_{\beta} + \tfrac{1}{2}\left(
3\hat{\mathcal{S}}^{\alpha}{}_{\beta}
+\hat{\Om}_{k}\,\d^{\alpha}{}_{\beta}
\right)\,\e^{-2t}
- \tfrac{1}{3}\,\hat{\Sig}^{\alpha}{}_{\beta}\,\e^{-3t}
+ \tfrac{1}{2}\,\hat{\Om}\,\d^{\alpha}{}_{\beta}
\,\e^{-3\gam t} + \bigO(\e^{-4t})\,\right] \ ,
\ee
with $\hat{E}^{\beta}{}_{i}$ and $\hat{E}_{\beta}{}^{i}$ related by
Eqs.~(\ref{hateinv}). It also follows from Eq.~(\ref{hq}) that the
Hubble scalar $H$ has the expansion
\be \lb{H_ae} H = \ell_{0}^{-1}\left[\,1 +
\tfrac{1}{2}\,\hat{\Om}_{k} \,\e^{-2t} +
\tfrac{1}{2}\,\hat{\Om}\,\e^{-3\gam t} + \bigO(\e^{-4t})\,\right]
\ . \ee
We then substitute Eqs.~(\ref{Ein_ae}) and (\ref{H_ae}) into
Eq.~(\ref{g_def}) to obtain the following expansion for the
3-metric~$g_{ij}$:
\be \lb{metric_ae} g_{ij} =  \ell_{0}^2\e^{2t}\,
\left[\,\hat{g}_{ij} + 3\hat{\mathcal{S}}_{ij}\,\e^{-2t} -
\tfrac{2}{3}\,\hat{\Sig}_{ij}\,\e^{-3t} + \bigO(\e^{-4t}) \right]
\ , \ee
where
\be \lb{gij_def} \hat{g}_{ij} =
\d_{\alpha\beta}\,\hat{E}^{\alpha}{}_{i}\,\hat{E}^{\beta}{}_{j} \
, \hspace{10mm} \hat{\mathcal{S}}_{ij} =
\hat{\mathcal{S}}_{\alpha\beta}\,\hat{E}^{\alpha}{}_{i}\,
\hat{E}^{\beta}{}_{j} \ , \hspace{10mm} \hat{\Sig}_{ij} =
\hat{\Sig}_{\alpha\beta}\,\hat{E}^{\alpha}{}_{i}\,
\hat{E}^{\beta}{}_{j} \ . \ee
Observe that the free functions $\hat{E}_{\alpha}{}^{i}$ determine
the coefficients $\hat{g}_{ij}$ in the leading order term in the
expansion through Eqs.~(\ref{hateinv}) and (\ref{gij_def}), and
also determine the second term $\hat{S}_{ij}$, while the free
functions $\hat{\Sig}_{ij}$ enter at a higher order.  Note that up
to order $\bigO(\e^{-4t})$, the matter coefficient $\hat{\Om}$ only
enters the spacetime metric (\ref{sepvol_metric}) through $H$ in
Eq.~(\ref{H_ae}).  It can be shown that the expansions (\ref{H_ae})
and (\ref{metric_ae}) are also valid for $\gam$ satisfying
$\tfrac43 \leq \gam < 2$, with the difference that the matter
coefficient enters into the $\bigO(\e^{-4t})$ term in
Eq.~(\ref{H_ae}).

Our results provide a confirmation of the ansatz for the metric
expansion given by Starobinski\v{\i}~\ct{sta83}, Eq.~(2). Some of
the details differ, however, due to the fact that Starobinski\v{\i}
employs the synchronous gauge.  We have established consistency
with his results by performing a coordinate transformation of the
form
\be
\lb{sync_to_sepvol}
\tilde{t}\equiv T = \int H^{-1}\,{\rm d}t + \dots \ , \hspace{10mm}
\tilde{x}^{i} = x^{i} + \dots \ .
\ee
We have also derived Starobinski\v{\i}'s metric expansion directly
in the synchronous gauge, using our integration method. 

The most striking feature of the expansion~(\ref{metric_ae}) is the
fact that the leading order coefficient in the expansion is an
arbitrary 3-metric, $\hat{g}_{ij}$, whereas the spacelike
3-surfaces ${\cal S}$:$\{t=\mbox{constant}\}$ in the de Sitter line
element~(\ref{ds_metric}) are intrinsically flat. The generality of
the leading order coefficient $\hat{g}_{ij}$ in the metric
expansion raises an apparent paradox --- a $G_{0}$ cosmology that
is future asymptotic to the spatially homogeneous and isotropic de
Sitter solution in the sense of the definition in
Subsec.~\ref{subsec:ds} can exhibit substantial spatial
inhomogeneity. We will show that this paradox is resolved by the
existence of event horizons.

%------------------------------
\subsubsection*{Event horizons}
%------------------------------
In cosmology an event horizon for a fundamental observer can be
thought of as constituted by the observer's past light cone at $t =
\infty$ (see Hawking and Ellis~\ct{hawell73}, p.~129). In order to
establish the existence of event horizons, we need to determine the
asymptotic form of geodesic null congruences. These are governed by
Eq.~(\ref{geo}), with the $K^{\alpha}$ determined by
Eqs.~(\ref{Keq})--(\ref{s}).

In App.~\ref{sec:eh_ae} we show that any null geodesic has the
asymptotic form
\be
\lb{eh_ae}
x^{i}(t) = x^{i}_{\infty}
- \hat{K}^{\alpha}\hat{E}_{\alpha}{}^{i}(x^{j}_{\infty})\,
\e^{-t} + \bigO(\e^{-2t}) \ ,
\ee
for constants $x^{i}_{\infty}$ and $\hat{K}^{\alpha}$. For given
$x^{i}_{\infty}$, the family of null geodesics then form an event
horizon for the fundamental observer whose worldline is $x^{i} =
x^{i}_{\infty}$.  The spatial distance from the observer to her/his
event horizon is given by
\be
d_{\rm H}(t) = \int_{0}^{1}
\sqrt{g_{ij}\,\frac{{\rm d}y^{i}}{{\rm d}s}\,
\frac{{\rm d}y^{j}}{{\rm d}s}}\,{\rm d}s \ ,
\ee
where $y^{i} = y^{i}(s)$ (with $0\leq s\leq1$) describes a
spacelike geodesic from $x^{i}_{\infty}$ to $x^{i}(t)$, for fixed
$t$.  In the asymptotic regime, the spacelike geodesic is basically
a straight line, and is approximated by the null geodesic
(\ref{eh_ae}) projected onto a spacelike 3-surface ${\cal
S}$:$\{t=\mbox{constant}\}$.  It follows from Eq.~(\ref{eh_ae})
that
\be
\lb{eh_dist}
d_{\rm H}(t) = \ell_{0} + \bigO(\e^{-t}) \ ,
\ee
where $\ell_{0}$ is given in Eq.~(\ref{ds_l0}).

One can always introduce local coordinates at $x^{i} =
x^{i}_{\infty}$ such that (see e.g., Schutz~\ct{sch85}, p.~156)
\be
\hat{g}_{ij}(x^{m}) = \ell_{0}^{2}\,\d_{ij}
+ \frac{\ptl^{2}\hat{g}_{ij}(x^{m}_{\infty})}{\ptl x^{k}\ptl x^{l}}
\,(x^{k}-x^{k}_{\infty})\,(x^{l}-x^{l}_{\infty}) + \dots \ .
\ee
For points within the event horizon the approximation
$\hat{g}_{ij}(x^{m}) \approx \ell_{0}^{2}\,\d_{ij}$ becomes
increasingly accurate as $t \rightarrow \infty$, due to
Eq.~(\ref{eh_ae}). In other words, {\em within\/} the event horizon
of a particular fundamental observer, the spacetime metric
asymptotically approaches the de Sitter metric.\footnote{Boucher
and Gibbons~\ct{bougib83} have clarified this matter in a different
way, by transforming the de Sitter metric to a static form, valid
within the event horizon of a particular fundamental observer.}
Since $\hat{g}_{ij}$ is general, however, the above approximation
cannot be done simultaneously at all points, reflecting the spatial
inhomogeneity of the $G_{0}$ cosmology at super-horizon scales. We
emphasize that the spatial inhomogeneity does not diminish as $t
\rightarrow \infty$ --- it is the observer who sees successively
smaller portions of that spatial inhomogeneity.

%-------------------------------------
\subsubsection*{Radiation bifurcation}
%-------------------------------------
A noteworthy feature of the asymptotic
expansion~(\ref{dS_bigO})--(\ref{dS_bigO_end}) is that for the
physically important case of incoherent radiation ($\gam =
\tfrac{4}{3}$), a bifurcation occurs that affects the peculiar
velocity $\vec{v}$ of the perfect fluid relative to the spacelike
3-surfaces ${\cal S}$:$\{t=\mbox{constant}\}$.

If $1 \leq \gam < \tfrac{4}{3}$, the components $v^{\alpha}$ tend
to zero, and as a result the fluid 4-velocity vector field
$\ti{\vecu}$ asymptotically coincides with the vorticity-free
timelike reference congruence $\vece_{0}$. For incoherent
radiation, the $v^{\alpha}$ do not tend to zero in general, and
$v^{2}$ can approach any value between 0 and 1. If $\tfrac{4}{3} <
\gam < 2$, $v^{2}$ tends to the value 1 as $t\rightarrow\infty$,
which means that the speed of the fluid relative to the fundamental
observers comoving with $\vece_{0}$ approaches the speed of light.

It is of considerable interest to investigate the limit behavior of
the fluid kinematical variables, as perceived by the fundamental
observers. Using the boost transformation laws provided in
App.~\ref{subsec:boost}, we find that the future asymptotic limits
for the Hubble-normalized fluid kinematical scalars (defined by
Eqs.~(\ref{fldlaccsc})--(\ref{fldlvorsc}) in
App.~\ref{subsec:boost}) are
\bea
\lb{dSfluscatilt1}
\lim_{t \rightarrow \infty}(\dot{\ti{U}}{}^{2},
\,\ti{\Sig}^{2}, \,\ti{W}^{2})
& = & (0, \,0, \,0) \hspace{10mm}
\text{for $1 \leq \gam < \tfrac{4}{3}$}. \\
\lim_{t \rightarrow \infty}(\dot{\ti{U}}{}^{2},
\,\ti{\Sig}^{2}, \,\ti{W}^{2})
& = & \left[\ \tfrac{1}{3}\,\hat{v}^2, \,0, \,0\ \right] \hspace{10mm}
\text{for $\gam = \tfrac{4}{3}$}. \\
\lb{dSfluscatilt2}
\lim_{t \rightarrow \infty}(\dot{\ti{U}}{}^{2},
\,\ti{\Sig}^{2}, \,\ti{W}^{2})
& = & \left[\ 3(\gam-1)^{2}, \,\tfrac{1}{4}\,(3\gam-4)^{2},
\,0\ \right] \hspace{10mm}
\text{for $\tfrac{4}{3} < \gam < 2$}.
\eea
This extends previous results given by Goliath and
Ellis~\ct{golell99} for some special SH cases with cosmological
constant to the general $G_{0}$ case (see also Goliath and
Nilsson~\ct{golnil2000} and Raychaudhuri and Modak
\ct{raymod88}). The context of spatial inhomogeneity stresses the
isotropization issue. We have to ask ourselves isotropization along
which timelike congruence? In the present context there is no
isotropization along the timelike reference congruence~$\vece_{0}$
when $\gam > \tfrac{4}{3}$, but this does not say anything about
whether the models isotropize along the fluid 4-velocity vector
field in this case, which, perhaps, is the physically best
motivated congruence to consider. However, to determine whether
this is the case or not takes us outside the scope of the present
paper, and so we leave this issue for future studies. We note that
when $\gam < \tfrac{4}{3}$, the fluid 4-velocity vector field
asymptotically coincides with the timelike reference congruence
and, hence, one does have isotropization in this case.

%%%%%%%%%%%%%%%%%%%%%%%%%%%%%%%%%%%%%%%%%%%%%%%%%%%%%%%%%%%%%%%%%%%
\section{Flat FL-like past asymptotics}
\lb{sec:flatfl}
%%%%%%%%%%%%%%%%%%%%%%%%%%%%%%%%%%%%%%%%%%%%%%%%%%%%%%%%%%%%%%%%%%%
In this section we give the detailed asymptotic form of the
Hubble-normalized variables for $G_{0}$ cosmologies that approach
the flat FL solution at early times.

%------------------------------------------------------------------
\subsection{$G_{0}$ cosmologies past asymptotic to the flat FL
  solution}
\lb{subsec:flatfl}
%------------------------------------------------------------------
The flat FL solution of Einstein's field equations describes a
cosmological model with perfect fluid matter source, whose density
parameter has the constant value $\Om = 1$.  The line element and
the fluid 4-velocity vector field are given by
\begin{gather}
\lb{flatfl_metric}
{\rm d}s^{2}
= -\,{\rm d}T^{2}
+ \ell_{0}^{2}\left(\frac{T}{T_{0}}\right)^{4/(3\gam)}\!
({\rm d}x^{2} + {\rm d}y^{2} + {\rm d}z^{2}) \ ,
\\
\ti{\vecu} = \ptl_{T} \ ,
\end{gather}
employing clock time $T$ as one of the local coordinates. The
matter energy density and pressure are
\be
\mu = \frac{4}{3\gam^{2}}\,T^{-2} \ , \hspace{10mm}
p = (\gam-1)\,\mu \ .
\ee
The Hubble scalar is
\be
\lb{naturalH}
H = \frac{2}{3\gam}\,T^{-1} \ ,
\ee
while the deceleration parameter is given by
\be
q = \tfrac{1}{2}(3\gam-2) \ ;
\ee
the latter is thus constant and positive. We find it convenient to
use a conformal time coordinate $\Eta$, defined by
\be
\lb{fl_eta}
\Eta := \Eta_{0}
\left(\frac{T}{T_{0}}\right)^{(3\gam-2)/3\gam} \ , \hspace{10mm}
\Eta_{0} := \frac{3\gam}{(3\gam-2)}\,\frac{T_{0}}{\ell_{0}} \ ,
\ee
with the unit of $\lgth$ $\ell_{0}$ and the constant $T_{0}$ chosen
to satisfy $\Eta_{0} = 1$, i.e.,
\be
\lb{fl_l0}
\ell_{0} = \frac{3\gam}{(3\gam-2)}\,T_{0} \ .
\ee
The line element is then cast into the form
\be {\rm d}s^{2} = \ell_{0}^{2}\,
\Eta^{4/(3\gam-2)}(-\,{\rm d}\Eta^{2} + {\rm d}x^{2}
+ {\rm d}y^{2} + {\rm d}z^{2}) \ ,
\ee
and the Hubble scalar is
\be
\lb{fl_H}
H = \frac{2}{3\gam}\,T_{0}^{-1}\,\Eta^{-3\gam/(3\gam-2)} \ .
\ee
We introduce the dynamical time coordinate $t$ via
\be
\left(\frac{T}{T_{0}}\right)^{2/(3\gam)} = \e^{t} \ ,
\ee
or, equivalently,
\be
\lb{fl_eta_t}
\Eta = \e^{\frac{1}{2}(3\gam-2)t} \ .
\ee
Relative to the natural orthonormal frame associated with the line
element~(\ref{flatfl_metric}), the Hubble-normalized variables are
[\,using Eq.~(\ref{fl_l0})\,]
\begin{gather}
\lb{flatfl_1}
E_{\alpha}{}^{i} = \tfrac{1}{2}\,(3\gam-2)\,
\e^{\frac{1}{2}(3\gam-2)t}\,\d_{\alpha}{}^{i} \\
\Sig_{\alpha\beta} = 0 \ , \hspace{10mm}
A^{\alpha} = 0 \ , \hspace{10mm}
N_{\alpha\beta} = 0 \ , \\
\lb{flatfl_3}
\Om = 1 \ , \hspace{10mm} \Oml = 0 \ .
\end{gather}
We note that the volume density $\cv$ is given by
\be
\cv = \ell_{0}^{3}\,\e^{3t}\ ,
\ee
so that the separable volume gauge conditions are satisfied.

Observe that the frame variables satisfy
$\lim_{t\rightarrow-\infty}E_{\alpha}{}^{i} = 0$. Thus, the flat FL
solution is described by an orbit in the Hubble-normalized state
space that is past asymptotic to an equilibrium point on the silent
boundary, $E_{\alpha}{}^{i} = 0$.

Motivated by Eqs.~(\ref{flatfl_1})--(\ref{flatfl_3}), we say that
{\em a $G_{0}$ cosmology is past asymptotic to the flat FL
solution\/} if the following limits are satisfied:
\begin{gather}
\lb{flatfl_c1}
\lim\limits_{t\rightarrow-\infty}(E_{\alpha}{}^{i}, r_{\alpha})^{T}
= \vec{0} \\
\lim\limits_{t\rightarrow-\infty}(\Sig_{\alpha\beta},
A^{\alpha}, N_{\alpha\beta})^{T} = \vec{0} \\
\lb{flatfl_c3}
\lim\limits_{t\rightarrow-\infty}\Om = 1 \ , \hspace{10mm}
\lim\limits_{t\rightarrow-\infty}\Oml = 0 \ .
\end{gather}
We note that for the flat FL solution the peculiar velocity
variables $v^{\alpha}$ are zero. We will show that if the limits
(\ref{flatfl_c1})--(\ref{flatfl_c3}) are satisfied, together with
certain technical restrictions, then the constraints imply that the
$v^{\alpha}$ tend to zero as $t\rightarrow-\infty$. We now consider
the class of $G_{0}$ cosmologies which are past asymptotic to the
flat FL solution, and present the asymptotic form of the
Hubble-normalized variables as $t \rightarrow -\infty$.

%------------------------------------------------------------------
\subsection{Asymptotic expansions}
\lb{subsec:flatfl_ae}
%------------------------------------------------------------------
In App.~\ref{app:flatfl}, we derive the asymptotic form of the
Hubble-normalized variables for $G_{0}$ cosmologies that are past
asymptotic to the flat FL solution, in the sense of
Eqs.~(\ref{flatfl_c1})--(\ref{flatfl_c3}). It is necessary, as in
Subsec.~\ref{subsec:ds_ae}, to impose certain restrictions on the
spatial derivatives; these are given in App.~\ref{app:flatfl}.

We now give the asymptotic expansions. For brevity we use $\Eta$
instead of $\e^{\frac{1}{2}(3\gam-2)t}$ [\,see
Eq.~(\ref{fl_eta_t})\,].
\bea
\lb{flatfl_bigO}
\left(E_{\alpha}{}^{i}, A^{\alpha}, N^{\alpha\beta}
\right)^{T}
& = & \left(\hat{E}_{\alpha}{}^{i},
\hat{A}^{\alpha}, \hat{N}^{\alpha\beta}\right)^{T} \Eta
+ {\cal O}(\Eta^{3}) \\
\lb{flatfldlr}
r_{\alpha} & = &
-\,\tfrac{1}{2}\,(\hat{E}_{\alpha}{}^{i}\,\ptl_{i}\hat{\Om}_{k})\,
\Eta^{3} + \bigO(\Eta^{5} + \Eta^{3+\frac{4}{(3\gam-2)}}) \\
\lb{flatfldlSig}
\Sig^{\alpha\beta}
& = & -\frac{6}{(3\gam+2)}\,\hat{\mathcal{S}}^{\alpha\beta}\,
\Eta^{2} + {\cal O}(\Eta^{4}) \\
\Oml & = & \frac{\Lambda}{3\hat{H}{}^{2}}\,
\Eta^{2+\frac{4}{(3\gam-2)}}\left[\,1
- \hat{\Om}_{k}\,\Eta^{2}
+ \bigO(\Eta^{4} + \Eta^{2+\frac{4}{(3\gam-2)}})\,\right]  \\
\lb{flatfldlom}
\Om & = & 1 - \hat{\Om}_{k}\,\Eta^{2}
+ {\cal O}(\Eta^{4} +\Eta^{2+\frac{4}{(3\gam-2)}}) \\
\lb{flatfl_bigO_end}
v^{\alpha} & = & \frac{1}{(3\gam+2)}\,\d^{\alpha\beta}\,
(\hat{E}_{\beta}{}^{i}\,\ptl_{i}\hat{\Om}_{k})\,\Eta^{3}
+ \bigO(\Eta^{5} + \Eta^{3+\frac{4}{(3\gam-2)}}) \ ,
\eea
where $\hat{H}$ is the leading order coefficient in the asymptotic
expansion of $H$:
\be
\lb{fl_H_ae}
H = \hat{H}\,\Eta^{-[1+\frac{2}{(3\gam-2)}]}\left[\,1
+ \tfrac{1}{2}\,\hat{\Om}_{k}\,\Eta^{2}
+ \bigO(\Eta^{4} + \Eta^{2+\frac{4}{(3\gam-2)}})\,\right] \ .
\ee
In the derivation in App.~\ref{app:flatfl}, we used the freedom to
re-define the 1-parameter family of spacelike 3-surfaces ${\cal
S}$:$\{t=\mbox{constant}\}$ while preserving the separable volume
gauge to set
\be
\lb{fl_Hhat}
\hat{H} = \frac2{3\gam}\,T_{0}^{-1} \ ,
\ee
so that $\hat{H} = \mbox{constant} > 0$. In addition,
Eq.~(\ref{ralpha}) and the constraints $({\cal C}_{\rm G})$ and
$({\cal C}_{\rm C})^{\alpha}$ were used in App.~\ref{app:flatfl} to
give the coefficients in Eqs.~(\ref{flatfldlr}), (\ref{flatfldlom})
and (\ref{flatfl_bigO_end}), respectively.\footnote{Note that up to
(and including) order $\Eta^{2}$, the present expansion also
satisfies the gauge conditions defining the constant mean curvature
($r_{\alpha} = 0$), synchronous ($\Udot_{\alpha} = 0$) and
fluid-comoving ($v^{\alpha} = 0$) temporal gauges, respectively.}

As in Subsec.~\ref{subsec:ds_ae} before, the coefficients
$\hat{A}^{\alpha}$, $\hat{N}_{\alpha\beta}$, $\hat{\Om}_{k}$ and
$\hat{\mathcal{S}}_{\alpha\beta}$ are determined by the
$\hat{E}_{\alpha}{}^{i}$ according to Eqs.~(\ref{Ahat}),
(\ref{Nhat}), (\ref{Omkhat}) and (\ref{Shat}). Likewise, the
3-Ricci curvature variables have the asymptotic expansions
\bea
\Om_{k} & = & \Eta^{2}\left[\,\hat{\Om}_{k} + \bigO(\Eta^{2})\,
\right] \\
{\cal S}_{\alpha\beta} & = & \Eta^{2}
\left[\,\hat{\cal S}_{\alpha\beta} + \bigO(\Eta^{2})\,\right] \ .
\eea
%

%------------------------------------------------------------------
\subsection{Features of asymptotic to flat FL $G_{0}$ cosmologies}
\lb{subsec:flatfl_is}
%------------------------------------------------------------------
In this section we investigate certain features of the class of
$G_{0}$ cosmologies that are past asymptotic to the flat FL
solution. One of our goals is to show that this class of
cosmologies admit an isotropic initial singularity in the sense of
Goode and Wainwright~\ct{goowai85} (see also Newman~\ct{new93a} and
Anguige and Tod~\ct{angtod99}).

%---------------------------------------------
\subsubsection*{Essential arbitrary functions}
%---------------------------------------------
Of the position-dependent coefficients in
Eqs.~(\ref{flatfl_bigO})--(\ref{flatfl_bigO_end}), only the nine
$\hat{E}_{\alpha}{}^{i}$ can be chosen arbitrarily, as suitably
differentiable functions of the $x^{i}$.  However, three of them
can be eliminated by the frame rotations~(\ref{rotation}), and
three by the spatial coordinate transformations~(\ref{x_trans}).
Thus, the asymptotic expansion contains~3 essentially arbitrary
functions of the spatial coordinates, as compared with~8 for a
general class of perfect fluid $G_{0}$ cosmologies. These models
thus form a set of measure zero in the Hubble-normalized state
space.

%--------------------------------
\subsubsection*{Metric expansion}
%--------------------------------
We now derive an asymptotic expansion for the spacetime metric in
order to relate our results to the work of other researchers, in
particular that of LK and Goode and Wainwright~\ct{goowai85}. We
find that the metric expansion sheds light on the nature of
large-scale spatial inhomogeneity in $G_{0}$ cosmologies that are
past asymptotic to the flat FL solution.

In the separable volume gauge, the line element has the form given
in Eq.~(\ref{sepvol_metric}). Equation~(\ref{dl13comts}) gives an
expansion for $\ptl_{t}E_{\alpha}{}^{i}$, which can be integrated
to give
\be
E_{\alpha}{}^{i} = \hat{E}_{\beta}{}^{i}\,\Eta \left[
\delta_{\alpha}{}^{\beta} + \left(
\frac{6}{(3\gam-2)(3\gam+2)}\,\hat{\mathcal{S}}_{\alpha}{}^{\beta}
-\,\tfrac{1}{2}\,\hat{\Om}_{k}\,\delta_{\alpha}{}^{\beta}
\right)\Eta^{2}
+ \bigO(\Eta^{4}+\Eta^{2+\frac{4}{(3\gam-2)}}) \right]\ .
\ee
Following the steps taken in Subsec.~\ref{subsec:ds_prop}, we
obtain the metric expansion
\be
\lb{fl_metric_ae}
g_{ij} = \Eta^{4/(3\gam-2)}\left[\,\hat{g}_{ij}
- \frac{12}{(3\gam-2)(3\gam+2)}\,\hat{\mathcal{S}}_{ij}\,\Eta^{2}
+ \bigO(\Eta^{4}+\Eta^{2+\frac{4}{(3\gam-2)}})\,\right] \ ,
\ee
where
\be
\hat{g}_{ij} = \hat{H}^{-2}\,\d_{\alpha\beta}\,
\hat{E}^{\alpha}{}_{i}\,\hat{E}^{\beta}{}_{j} \ ,
\hspace{10mm}
\hat{\mathcal{S}}_{ij} = \hat{H}^{-2}\,
\hat{\mathcal{S}}_{\alpha\beta}\,\hat{E}^{\alpha}{}_{i}\,
\hat{E}^{\beta}{}_{j} \ .
\ee
Our result provides a rigorous confirmation of the ansatz for the
metric expansion made by LK, Eq.~(4.1), for the case of incoherent
radiation. Some of the details differ, however, due to the fact
that LK employ the synchronous gauge.  We have established
consistency with their results by performing a coordinate
transformation of the form given in Eqs.~(\ref{sync_to_sepvol}).
We have also derived LK's metric expansion directly in the
synchronous gauge, using our integration method.

%-----------------------------------------------
\subsubsection*{Isotropic initial singularities}
%-----------------------------------------------
We briefly digress to introduce the notion of an isotropic initial
singularity; see Ref.~\ct{goowai85}. A cosmological
model\footnote{We assume that the matter source is a perfect fluid
satisfying Eqs.~(\ref{eos}) and~(\ref{gam}).}  is said to admit an
{\em isotropic initial singularity\/} if the physical spacetime
metric $\ten{g}$ is conformal to an unphysical
metric~$\tilde{\ten{g}}$:
\be
\ten{g} = \chi^{2}\,\ti{\ten{g}} \ ,
\ee
where $\chi = \chi(\Eta)$ is a differentiable function of a time
coordinate $\Eta$ and satisfies $\chi(0) = 0$, while the conformal
metric $\ti{\ten{g}}$ is regular on the spacelike 3-surface
$\ti{{\cal S}}$:$\{\Eta=0\}$. The conformal factor $\chi(\Eta)$ is
also required to satisfy
\begin{align}
\lb{4a}
\lim_{\Eta\rightarrow 0^{+}}\frac{\chi^{\prime}(\Eta)}{\chi(\Eta)}
&= \infty \\
\lb{4b}
\lim_{\Eta\rightarrow 0^{+}}
\frac{\chi^{\prime\prime}(\Eta)}{\chi(\Eta)}
\left(\frac{\chi^{\prime}(\Eta)}{\chi(\Eta)}\right)^{-2}
&= -\,\tfrac{1}{2}\,(3\gam-4)\ .
\end{align}
It follows from Eqs.~(\ref{sepvol_metric}), (\ref{fl_H_ae}) and
(\ref{fl_metric_ae}) that the spacetime metric for the class of
$G_{0}$ cosmologies that are past asymptotic to the flat FL
solution does satisfy Eqs.~(\ref{4a}) and (\ref{4b}) with
\be
\chi = \ell_{0}\,\Eta^{2/(3\gam-2)} \ ,
\ee
where $\Eta$ is the conformal time coordinate.

%---------------------------------
\subsubsection*{Particle horizons}
%---------------------------------
The generality of the leading order coefficient $\hat{g}_{ij}$ in
the metric expansion (\ref{fl_metric_ae}), which should be compared
with Eq.~(\ref{metric_ae}), raises an apparent paradox --- a
$G_{0}$ cosmology that is past asymptotic to the spatially
homogeneous and isotropic flat FL solution in the sense of the
definition in Subsec.~\ref{subsec:flatfl} can exhibit substantial
spatial inhomogeneity. We will show that this paradox is resolved
by the existence of particle horizons.

In order to establish the existence of particle horizons, we need
to determine the asymptotic form of geodesic null
congruences. These are governed by Eq.~(\ref{geo}), with the
$K^{\alpha}$ determined by Eqs.~(\ref{Keq})--(\ref{s}).

In App.~\ref{sec:ph_ae} we show that any past-oriented null geodesic has
the following asymptotic form as $\Eta \rightarrow 0$:
\be
\lb{ph_ae}
x^{i}(\Eta) = x^{i}_{\rm BB} +\frac{2}{(3\gam-2)}\,\hat{K}^{\alpha}
\hat{E}_{\alpha}{}^{i}(x^{j}_{\rm BB})\,\Eta
+ \bigO(\Eta^2) \ ,
\ee
for constants $x^{i}_{\rm BB}$ and $\hat{K}^{\alpha}$.  The
constants $x^{i}_{\rm BB}$ give the point of termination of the
null geodesic at the singularity $\Eta=0$.  Thus the past null cone
at time $\Eta$ for the fundamental observer whose worldline is
$x^{i} = x^{i}_{0}$ will intersect the singularity $\Eta=0$,
thereby defining a particle horizon, which is formed by the
fundamental worldlines given by $x^{i} = x^{i}_{\rm BB}$.  The
spatial distance from the observer to her/his particle horizon at
time $\Eta$ is given by
\be
d_{\rm H}(\Eta) = \int_{0}^{1}
\sqrt{g_{ij}\,\frac{{\rm d}y^{i}}{{\rm d}s}\,
\frac{{\rm d}y^{j}}{{\rm d}s}}\,{\rm d}s \ ,
\ee
where $y^{i} = y^{i}(s)$ (with $0\leq s\leq1$) describes a
spacelike geodesic from $x^{i}_{\rm BB}$ to $x^{i}(\Eta)$, for
fixed $\Eta$.  As in the derivation of Eq.~(\ref{eh_dist}), it
follows from Eq.~(\ref{ph_ae}) that
\be
d_{\rm H}(\Eta)= \Eta^{1+\frac{2}{(3\gam-2)}}
\left[\,\ell_{0} + \bigO(\Eta)\,\right] \ ,
\ee
where $\ell_{0}$ is given in Eq.~(\ref{fl_l0}).

One can introduce local coordinates at $x^{i} = x^{i}_{\rm BB}$
such that
\be
\hat{g}_{ij}(x^{m}) = \ell_{0}^{2}\,\d_{ij}
+ \frac{\ptl^{2}\hat{g}_{ij}(x^{m}_{\rm BB})}{\ptl x^{k}\ptl x^{l}}
\,(x^{k}-x^{k}_{\rm BB})\,(x^{l}-x^{l}_{\rm BB}) + \dots \ .
\ee
For points within the particle horizon the approximation
$\hat{g}_{ij}(x^{m}) \approx \ell_{0}^{2}\,\d_{ij}$ becomes
increasingly accurate as $\Eta \rightarrow 0$, due to
Eq.~(\ref{ph_ae}). In other words, {\em within\/} the particle
horizon of a particular fundamental observer, the spacetime metric
is close to the flat FL metric. Since $\hat{g}_{ij}$ is general,
however, the above approximation cannot be done simultaneously at
all points, reflecting the spatial inhomogeneity of the $G_{0}$
cosmology at super-horizon scales. We emphasize that the spatial
inhomogeneity does not diminish --- it is the observer who sees
successively smaller portions of that spatial inhomogeneity.

%%%%%%%%%%%%%%%%%%%%%%%%%%%%%%%%%%%%%%%%%%%%%%%%%%%%%%%%%%%%%%%%%%%
\section{Concluding remarks}
\lb{sec:disc}
%%%%%%%%%%%%%%%%%%%%%%%%%%%%%%%%%%%%%%%%%%%%%%%%%%%%%%%%%%%%%%%%%%%
In this paper, we have given a unified analysis, within the
framework of the Hubble-normalized state space, of $G_{0}$
cosmologies that undergo asymptotic isotropization, either at late
times (asymptotic to the de Sitter solution, with $\ell \rightarrow
\infty$) or near the initial singularity (asymptotic to the flat FL
solution, with $\ell\rightarrow 0$).

The analysis reveals a number of common features as regards the
asymptotic behavior in the two classes, as well as a number of
significant differences, which emerge most clearly if we write the
asymptotic expansions in Subsec~\ref{subsec:ds_ae} in terms of the
conformal time coordinate of the de Sitter line element, given by
Eq.~(\ref{ds_eta}).  We note that the length scale factor is given
by
\be
\ell(\Eta) = \ell_{0}\,\Eta^{-1} \ ,
\ee
for the de Sitter solution, and by
\be
\ell(\Eta) = \ell_{0}\,\Eta^{2/(3\gam-2)} \ ,
\ee
for the flat FL solution. Thus, for both asymptotes $\Eta
\rightarrow 0^+$ in the asymptotic regimes under consideration.

Firstly, the shear rate $\Sig_{\alpha\beta}$ and anisotropic
3-Ricci curvature $\mathcal{S}_{\alpha\beta}$ have the same leading
asymptotic dependence on $\Eta$ as $\Eta \rightarrow 0^+$:
\begin{align}
\lb{sig_common}
\Sig_{\alpha\beta} &= C\,\hat{\mathcal{S}}_{\alpha\beta}\,\Eta^{2}
+ \bigO(\Eta^{3+\delta}) \\
\mathcal{S}_{\alpha\beta} &= \hat{\mathcal{S}}_{\alpha\beta}\,
\Eta^{2} + \bigO(\Eta^{3+\delta}) \ ,
\end{align}
where $\delta \geq 0$, and $C = -\,3$ in the de Sitter case while
$C = -\,6/(3\gam+2)$ in the flat FL case.

Secondly, there is a common asymptotic form for the line element,
namely
\be
\lb{asymp_metric}
{\rm d}s^{2} = \ell^{2}(\Eta)\left[\,-\,{\rm d}\Eta^{2}
+ \ell_{0}^{-2}\,\hat{g}_{ij}\,{\rm d}x^{i}\,{\rm d}x^{j}
+ \bigO(\Eta^{2})\,\right] \ ,
\ee
as $\Eta \rightarrow 0^+$, where $\ell(\Eta)$ is the length scale
factor for the de Sitter solution or for the flat FL solution.
Likewise, the Hubble scalar has the common form
\be
H = H_{\rm asymp}(\Eta)\left[\,1 + \tfrac{1}{2}\,\hat{\Om}_{k}\,
\Eta^{2} + \bigO(\Eta^{3+\delta})\,\right] \ ,
\ee
where $\delta \geq 0$, and $H_{\rm asymp}(\Eta)$ is the Hubble
scalar for the de Sitter solution or for the flat FL solution,
respectively [\,see Eqs.~(\ref{ds_H}), (\ref{H_ae}), (\ref{fl_H}),
(\ref{fl_H_ae}) and (\ref{fl_Hhat})\,]. Here $\hat{\Om}_{k}$ is the
leading order coefficient in the expansion for the 3-Ricci
curvature scalar $\Om_{k}$, which has a common form for the two
classes, namely,
\be
\Om_{k} = \hat{\Om}_{k}\,\Eta^{2} + \bigO(\Eta^{4}) \ .
\ee
The key feature of the asymptotic line element (\ref{asymp_metric})
is the presence of the arbitrary 3-metric $\hat{g}_{ij}(x^{k})$.
We have shown that the spatial inhomogeneity that it generates is
significant only at super-horizon scales (the event horizon for
$G_{0}$ cosmologies that are future asymptotic to the de Sitter
solution and the particle horizon for $G_{0}$ cosmologies that are
past asymptotic to the flat FL solution).

The common asymptotic features of the two classes can be understood
to some extent within the context of the Hubble-normalized state
space.  For each class, the asymptotic state is described by an
equilibrium point on the silent boundary,
$$
E_{\alpha}{}^{i} = 0 \ ,
$$
[\,see~Paper~I, Eq.~(75)\,], which is associated with an isotropic
and conformally flat solution of Einstein's field equations.  Use
of a conformal time coordinate $\Eta$, which tends to zero on the
silent boundary for both classes, leads to common asymptotic decay
rates for its various Hubble-normalized quantities.  The
asymptotically silent dynamics ($E_{\alpha}{}^{i} \rightarrow 0$)
is also responsible for the existence of an event horizon and a
particle horizon, as shown in Subsecs.~\ref{subsec:ds_ae}
and~\ref{subsec:flatfl_ae}.

The differences between the two classes originate in the asymptotic
behavior of the deceleration parameter $q$, which is negative in
the de Sitter case ($q \rightarrow -1$) and positive in the flat FL
case ($q\rightarrow \frac12(3\gam-2)$).  This difference affects
the higher-order terms in the asymptotic expansion of the shear
rate~$\Sig_{\alpha\beta}$. In the de Sitter case, there is a term
$\hat{\Sig}_{\alpha\beta}\,\Eta^{3}$, where the
$\hat{\Sig}_{\alpha\beta}$ are freely specifiable functions, while
in the flat FL case, this term does not appear.\footnote{The second
shear rate mode in this case is a growing mode (into the past), and
the assumption of the isotropization requires the corresponding
coefficient to be zero.} The absence of the
$\hat{\Sig}_{\alpha\beta}$-term, of course, accounts for the
difference in the number of freely specifiable functions in the two
cases. This difference is also reflected in the asymptotic
expansions for the Hubble-normalized electric and magnetic parts of
the Weyl curvature [\,see Paper~I, Eqs.~(152) and~(153)\,]. For the
electric Weyl curvature we have in the de Sitter case
\be
{\cal E}_{\alpha\beta} = \tfrac{1}{3}\,\hat{\Sig}_{\alpha\beta}\,
\Eta^{3} + \bigO(\Eta^{4}) \ ,
\ee
and in the flat FL case,
\be
{\cal E}_{\alpha\beta} = \frac{3\gam}{(3\gam+2)}\,
\hat{\mathcal{S}}_{\alpha\beta}\,\Eta^{2} + \bigO(\Eta^{4}) \ .
\ee
For the magnetic Weyl curvature we have in both cases
\be
\lb{mag_common}
{\cal H}_{\alpha\beta} = -\,\hat{\cal C}_{\alpha\beta}\,
\Eta^{3} + \bigO(\Eta^{4}) \ ,
\ee
where $\hat{\cal C}_{\alpha\beta}$ is given by the ``hatted"
version of Eq.~(\ref{dl3ct}), with $r_{\alpha}$ set to zero.

The difference between the de Sitter and the flat FL cases as
regards the number of freely specifiable functions can be
interpreted in a dynamical systems context.  We have already noted
that both the de Sitter solution~(\ref{ds_metric}) and the flat FL
solution~(\ref{flatfl_metric}) determine equilibrium points on the
silent boundary in the Hubble-normalized state space.  It follows
that the orbits of $G_{0}$ cosmologies that are future asymptotic
to the de Sitter solution (respectively, past asymptotic to the
flat FL solution) are asymptotic to these equilibrium points.
The asymptotic expansion contains 8 arbitrary functions (the
maximum possible number) in the de Sitter case, which confirms that the
de Sitter equilibrium point is asymptotically stable, as proved in
App.~\ref{app:ds} (see Stage 0). On the other
hand, the presence of only 3 arbitrary functions in the flat FL
case implies that the flat FL equilibrium point has both a stable
and an unstable manifold (into the past), and our asymptotic
solutions describe the stable manifold.

A major accomplishment of the present paper is to provide
derivations of the detailed asymptotic properties of $G_{0}$
cosmologies that undergo asymptotic isotropization, instead of
simply making {\em ad hoc} assumptions about the form of the
metric, as has been done previously.  Our derivations do, however,
depend in a significant way on the assumption that the spatial
partial derivatives of the Hubble-normalized variables remain
bounded in the asymptotic regimes.  It is this assumption that
enables us to rely exclusively on analytic methods for systems of
ODE.

It would certainly be desirable to attempt to use methods from the
theory of PDE to prove the validity of the assumption of bounded
spatial derivatives for cosmological models that undergo asymptotic
isotropization.  Recent experience with $G_{2}$ cosmologies, which
we now describe, gives some cause for optimism.  Firstly, it is
known that in $G_{2}$ cosmologies spatial partial derivatives can
diverge on approach to the initial singularity, through the
creation of so-called Gowdy spikes (see Paper~I, Subsec.~4.1, and
Ref.~\ct{renwea2001}). This spatial structure is created, however,
by the local instability of the Kasner solutions. Since these
solutions do not play a r\^{o}le in the present context (the
approach to an isotropic initial singularity), we do not expect
spatial structure of this nature to develop.  Secondly, at late
times, support for our assumption is provided by numerical
simulations\footnote{Woei Chet Lim, unpublished.}  of $G_{2}$
cosmologies, which do not show the development of large spatial
derivatives.

A second limitation of our analysis is that it is local in nature
in that we restrict our considerations to some open subset of the
integral curves of the timelike reference congruence~$\vece_{0}$.
This restriction is inevitable, however, since in a $G_{0}$
cosmology all timelines do not necessarily share the same
asymptotic evolution.  For example, a $G_{0}$ cosmology with
positive cosmological constant may approach the de Sitter solution
along some timelines but may undergo collapse to a future
singularity along others.

Another matter that requires comment is the choice of temporal
gauge.  In the present paper we have chosen to work with the {\em
separable volume gauge\/}, defined by Eqs.~(\ref{sepvol}), because
we have found it to be a convenient computational gauge.  On the
other hand, we have been able to transform our asymptotic
expansions to the synchronous gauge, and find that the leading
order terms, e.g., Eqs.~(\ref{sig_common})--(\ref{mag_common}), are
unchanged. This calculation thus provides evidence for the {\em
gauge robustness\/} of our results.

In conclusion, it would be of interest to relate our results to
perturbation analyses of the de Sitter solution (see for example
Barrow~\ct{bar83} and Bruni {\em et al\/}~\ct{bruetal2002}) and of
the flat FL solution. We have not been able to clarify the
relationship to our satisfaction, particularly in the case of the
de Sitter solution, and so we leave this matter for future
investigation.  In addition, other possible applications of the
Hubble-normalized state space framework for $G_{0}$ cosmologies
presented here that suggest themselves naturally are establishing
firm links with mainstream issues in observational cosmology such
as the cosmic background radiation, peculiar motions of galaxies,
or gravitational lensing.

%%%%%%%%%%%%%%%%%%%%%%%%%%%%%%%%%%%%%%%%%%%%%%%%%%%%%%%%%%%%%%%%%%%
\section*{Acknowledgments}
%%%%%%%%%%%%%%%%%%%%%%%%%%%%%%%%%%%%%%%%%%%%%%%%%%%%%%%%%%%%%%%%%%%
H.v.E. acknowledges repeated kind hospitality by the Department of
Physics, University of Karlstad, Sweden. C.U. was in part
supported by the Swedish Research Council. J.W. was in part
supported by a grant from the Natural Sciences and Engineering
Research Council of Canada.

\appendix
%%%%%%%%%%%%%%%%%%%%%%%%%%%%%%%%%%%%%%%%%%%%%%%%%%%%%%%%%%%%%%%%%%%
\section{Asymptotic results for ordinary differential equations}
\lb{app:prop}
%%%%%%%%%%%%%%%%%%%%%%%%%%%%%%%%%%%%%%%%%%%%%%%%%%%%%%%%%%%%%%%%%%%

%-----------------------------
\subsubsection*{Proposition 1}
%-----------------------------
Given
\be
\lb{lem1_1}
\ptl_{t}M_{\alpha\beta} = A_{\alpha}{}^{\gam}M_{\gam\beta}
+ B_{\alpha\beta} \ ,
\ee
where $M_{\alpha\beta}$, $A_{\alpha\beta}$ and $B_{\alpha\beta}$
are Cartesian tensors.\footnote{We will also need a vector version
of this result, i.e., the DE is of the form
$$
\ptl_{t}M^{\alpha} = A^{\alpha}{}_{\beta}M^{\beta} + B^{\alpha} \ .
$$
}
If
\be
\lb{lem1_2}
\lim_{t\rightarrow\infty} A_{\alpha\beta}(t)
= -k\delta_{\alpha\beta} \ , \hspace{10mm} k > 0 \ ,
\ee
and
\be
\lb{lem1_3}
B_{\alpha\beta}=\bigO(\e^{-lt}) \ , \hspace{10mm} l > 0 \ ,
\ee
then for any given $\eps > 0$,
\be
M_{\alpha\beta} = \bigO(e^{(-m+\eps)t}) \ ,
\ee
where $m=\min(k,l)$.

\paragraph{Proof:}

Introduce
\be
M = \left(M_{\alpha\beta}M^{\alpha\beta}\right)^{1/2} \ , \hspace{10mm}
M_{\alpha\beta} = M S_{\alpha\beta} \ , \hsp5 \text{where }
\left(S_{\alpha\beta}S^{\alpha\beta}\right)^{1/2} = 1 \ .
\ee
It follows from Eqs.~(\ref{lem1_1}) that
\be
\lb{proof1_2}
\ptl_{t}M = \left(S^{\alpha\beta}A_{\alpha}{}^{\gam}S_{\gam\beta}\right)M
+ S^{\alpha\beta}B_{\alpha\beta} \ .
\ee
Now Eq.~(\ref{lem1_2}) implies that for all $\eps > 0$ there exists
a $t_{0}$ such that
\be
\lb{proof1_3}
|S^{\alpha\beta}(A_{\alpha}{}^{\gam} + k\d_{\alpha}{}^{\gam}
S_{\gam\beta}| \leq \eps \ ,
\ee
for all $t \geq t_{0}$, and Eq.~(\ref{lem1_3}) implies there exists
a $C > 0$ such that
\be
\lb{proof1_4}
|S^{\alpha\beta}B_{\alpha\beta}| \leq C \e^{-lt} \ .
\ee
It then follows that
\be
\ptl_{t}M \leq (-k+\eps)M + C\e^{-lt} \ .
\ee
Multiplying by $\e^{(k-\eps)t}$ and integrating from $t = t_{0}$ to
$t$, yields
\be
M \leq \left[M_{0} + \frac{C\e^{-lt_0}}{(-k+\eps+l)}\right]
\e^{(-k+\eps)(t-t_0)}
+ \frac{C\e^{-lt_0}}{(k-\eps-l)}\e^{-l(t-t_0)} \ ,
\ee
from which the result follows.  \hfill $\Box$

%-----------------------------
\subsubsection*{Proposition 2}
%-----------------------------
Given
\be
\lb{lem2_1}
\ptl_{t}M_{\alpha\beta} = A_{\alpha}{}^{\gam}M_{\gam\beta}
+ B_{\alpha\beta} \ ,
\ee
where $M_{\alpha\beta}$, $A_{\alpha\beta}$ and $B_{\alpha\beta}$
are Cartesian tensors. If
\be
\lb{lem2_2}
\lim_{t\rightarrow\infty} A_{\alpha\beta}(t) = k\d_{\alpha\beta}
\ , \hspace{10mm} k > 0 \ ,
\ee
\be
B_{\alpha\beta} = \bigO(\e^{-lt}) \ , \hspace{10mm} l > 0 \ ,
\ee
and $M_{\alpha\beta}$ is bounded, then
\be
M_{\alpha\beta} = \bigO(e^{-lt}) \ .
\ee

\paragraph{Proof:}

As in the proof of Proposition~1, we obtain Eqs.~(\ref{proof1_2})
and~Eq.~(\ref{proof1_3}) with $k$ replaced by $-k$, and
Eq.~(\ref{proof1_4}). It follows from these equations that
\be
\ptl_{t}M \geq M (k-\eps) - C \e^{-lt} \ .
\ee
Multiply by $\e^{-(k-\eps)t}$ and integrate from $t$ to $\infty$,
knowing that $M$ is bounded, to obtain
\be
M \leq \frac{C}{(k-\eps+l)} \e^{-lt} \ ,
\ee
and the result follows. \hfill $\Box$

%-----------------------------
\subsubsection*{Proposition 3}
%-----------------------------
Given
\be
\ptl_{t}\mat{M} = -\,k\mat{M} + \mat{B} \ ,
\ee
where $\mat{M}$ and $\mat{B}$ are $m \times n$ matrices and $k>0$
is constant. If $\mat{B} = \bigO(\e^{-lt})$, $l > k$, then
\be
\mat{M} = \hat{\mat{M}}\,\e^{-kt} + \bigO(\e^{-lt}) \ .
\ee

\paragraph{Proof:} Rewrite the differential equation (DE) as
$\ptl_{t}(\e^{kt}\mat{M})=\e^{kt}\mat{B}$, and integrate. \hfill
$\Box$

%%%%%%%%%%%%%%%%%%%%%%%%%%%%%%%%%%%%%%%%%%%%%%%%%%%%%%%%%%%%%%%%%%%
\section{Derivation of asymptotically de Sitter future dynamics}
\lb{app:ds}
%%%%%%%%%%%%%%%%%%%%%%%%%%%%%%%%%%%%%%%%%%%%%%%%%%%%%%%%%%%%%%%%%%%
In this appendix we derive the asymptotic
expansions~(\ref{dS_bigO})--(\ref{dS_bigO_end}) for $G_{0}$
cosmologies that are future asymptotic to the de Sitter solution in
the sense of Eqs.~(\ref{ds_c1})--(\ref{ds_c3}).  For ease of
reference, we list the complete set of restrictions below.

\begin{enumerate}[{A}1:]
\item $E_{\alpha}{}^{i},\ A^{\alpha},\ N_{\alpha\beta},\
    r_{\alpha},\ \Sig_{\alpha\beta},\ \Om$ and $\Oml-1$ are
    sufficiently small at some initial time which we can take to be
    $t = 0$.
\item The partial derivatives of any dimensionless variable $V$ are
    bounded as $t\rightarrow\infty$. In addition, if $V$ tends to
    zero, then $\ptl_{i}V = \bigO(\| V \|)$ as $t \rightarrow
    \infty$, for fixed $x^{i}$, where $\| V \|$ is the magnitude of
    $V$.\footnote{For example, $\|E_{\alpha}{}^{i}\|=
    \sqrt{\d^{\alpha\beta}E_{\alpha}{}^{i}E_{\beta}{}^{i}}$, for
    each $i$, and $\|\Sig_{\alpha\beta}\| =\Sig$.}
\item If $V = \hat{V}\,\e^{-nt} + \bigO(\e^{-mt})$ as $t
    \rightarrow \infty$, where $m > n > 0$, then $\ptl_{i}V =
    \ptl_{i}\hat{V}\,e^{-nt} + \bigO(e^{-mt})$ as $t \rightarrow
    \infty$.
\end{enumerate}

%-----------------------
\subsubsection*{Stage 0}
%-----------------------
We first establish asymptotic stability of the de Sitter solution.
The evolution equations for the variables in A1 are of the form
\be
\ptl_{t}\vec{X} = \mat{A}\vec{X} + \vec{f}(t,\vec{X}) \ ,
\ee
where $\vec{X}$ is the vector containing the variables in A1,
$\mat{A}$ is a constant diagonal real-valued matrix with negative
eigenvalues, while $\vec{f}(t,\vec{X})$ contains products of components
in $\vec{X}$ and $\parb_{\alpha}\vec{X}$. Assumptions A1 and A2 imply
that $\vec{f}(t,\vec{X})$ is continuous for all $t \geq 0$ and
$\vec{f}(t,\vec{X}) = o(\|\vec{X}\|)$ as $\|\vec{X}\| \rightarrow
0$. Then Theorem~1.1 in Coddington and Levinson~\ct{codlev55}, p.~314,
implies that $\vec{X} = \vec{0}$ is asymptotically stable. Thus, we
obtain the property
\begin{enumerate}[{P}1:]
\item $E_{\alpha}{}^{i},\ A^{\alpha},\ N_{\alpha\beta},\
      r_{\alpha},\ \Sig_{\alpha\beta},\ \Om,\ \Oml-1 \rightarrow 0$
      as $t \rightarrow \infty$, for fixed $x^{i}$.
\end{enumerate}
We organize the derivation of asymptotic expansions into two
stages, in which we repeatedly use the evolution equations and the
constraints $({\cal C}_{\rm G})$ and $({\cal
C}_{\Lambda})_{\alpha}$. In Stage~1, we obtain order estimates for
all variables, primarily using Propositions~1 and 2 in
App.~\ref{app:prop}. In Stage~2, we use the results from Stage~1 in
conjunction with Proposition~3 to obtain the explicit asymptotic
expansions. The assumptions A2 and A3 are needed throughout in
order to bound the spatial derivative terms in the evolution
equations and in the constraints $({\cal C}_{\rm G})$ and~$({\cal
C}_{\Lambda})_{\alpha}$.

%-----------------------
\subsubsection*{Stage 1}
%-----------------------
The deceleration parameter $q$, given by Eq.~(\ref{hdecel}), plays
a dominant r\^{o}le in the evolution equations. Its limiting value
as $t\rightarrow\infty$ follows from P1 and A2:
\be
\lb{ds_r1_q}
\lim_{t\rightarrow\infty}q = -\,1 \ .
\ee
We begin by considering $E_{\alpha}{}^{i}$. Using P1 and
Eq.~(\ref{ds_r1_q}), the evolution equation~(\ref{dl13comts}) for
$E_{\alpha}{}^{i}$ assumes the form of Eqs.~(\ref{lem1_1}) and
(\ref{lem1_2}) with $k = 1$ and $B_{\alpha\beta} = 0$. It follows
from Proposition~1 that for any given $\eps_{E} > 0$,
\be
\lb{ds_r1_E}
E_{\alpha}{}^{i} = \bigO(\e^{(-1+\eps_{E})t}) \ ,
\ee
as $t \rightarrow \infty$.

We can now use Eqs.~(\ref{ds_r1_q}), (\ref{ds_r1_E}) and P1, A2
to conclude that the evolution equation~(\ref{dladot}) for
$A^{\alpha}$ assumes the form of Eqs.~(\ref{lem1_1}) and
(\ref{lem1_2}) with $k = 1$ and $l = 1-\eps_{E}$. It follows from
Proposition~1 that for any given $\eps_{A} > \eps_{E}$,
\be
\lb{ds_r1_A}
A^{\alpha} = \bigO(\e^{(-1+\eps_{A})t}) \ ,
\ee
as $t \rightarrow \infty$.

At this stage we have two different epsilons in
Eqs.~(\ref{ds_r1_E}) and~(\ref{ds_r1_A}), with $\eps_{A} >
\eps_{E}$. Equation~(\ref{ds_r1_E}) is equally valid, however, if
we replace $\eps_{E}$ by $\eps_{A}$, and in the interests of
simplicity we choose to do this, dropping the subscripts on the
epsilons. We will also make this $\eps$-simplification in
subsequent steps.

In a similar way, Eq.~(\ref{dlndot}) leads to\footnote{Here and in
the future, we will omit the qualifier $t\rightarrow\infty$.}
\be
\lb{ds_r1_N}
N^{\alpha\beta} = \bigO(\e^{(-1+\eps)t}) \ .
\ee

Next, we derive intermediate results for $\Om$ and $r_{\alpha}$.
For any given $\delta > 0$, P1 and A2 imply that\footnote{Here we
use A2 to write $\ptl_{i}\Om = \bigO(\Om)$, and also use the fact
that $E_{\alpha}{}^{i} \rightarrow 0$ to write $|E_{\alpha}{}^{i}|
< \eps$, for $t$ sufficiently large.\lb{foot1}}
\be
\lb{ds_r1_pom}
-\,\frac{\gam}{G_{+}}\,v^{\alpha}\,\parb_{\alpha}\Om
\leq \delta\,\Om \ ,
\ee
for $t$ sufficiently large. Then, for any given $\eps > 0$,
Eqs.~(\ref{dlomdot}) and (\ref{ds_r1_q})--(\ref{ds_r1_pom}) give
\be
\lb{ds_Om_1}
\ptl_{t}\Om \leq \left[\,-\,\frac{\gam}{G_{+}}\,(3+v^{2})
+ \eps\,\right]\Om \ ,
\ee
for $t$ sufficiently large. The inequalities $1 \leq \gam < 2$
imply that
\be
\lb{ds_Om_2}
-\,\frac{\gam}{G_{+}}\,(3+v^{2}) \leq -\,3 \ .
\ee
It follows from Eqs.~(\ref{ds_Om_1}) and (\ref{ds_Om_2}) by
integrating that
\be
\lb{ds_Om_3}
\Om = \bigO(\e^{(-3+\eps)t}) \ .
\ee

The constraint $({\cal C}_{\Lambda})_{\alpha}$, in conjunction with
P1, A2 and Eq.~(\ref{ds_r1_E}), implies
\be
\lb{ds_r_1}
r_{\alpha} = \bigO(\e^{(-1+\eps)t}) \ .
\ee
With Eqs.~(\ref{ds_Om_3}) and (\ref{ds_r_1}), we now have enough
information to derive an asymptotic expression for
$\Sig^{\alpha\beta}$. Indeed, Eq.~(\ref{dlsigdot}) assumes the form
of Eqs.~(\ref{lem1_1}) and (\ref{lem1_2}) with $k = 3$ and $l =
2-\eps$, which implies, using Proposition~1, that
\be
\lb{ds_r1_Sig}
\Sig^{\alpha\beta} = \bigO(\e^{(-2+\eps)t}) \ .
\ee
It follows from Eqs.~(\ref{ds_r1_E})--(\ref{ds_r1_N}),
(\ref{ds_r_1}) and A2 that $\Om_k = \bigO(\e^{(-2+\eps)t})$, which
in turn implies, using Eqs.~(\ref{ds_Om_3}), (\ref{ds_r1_Sig}) and
the constraint $({\cal C}_{\rm G})$, that
\be
\Oml-1 = \bigO(\e^{(-2+\eps)t}) \ .
\ee
By using A2 and the constraint $({\cal C}_{\Lambda})_{\alpha}$, we
can now conclude that
\be
r_{\alpha} = \bigO(\e^{(-3+\eps)t}) \ .
\ee

It remains to derive the decay rates of $\Om$ and $v^{\alpha}$.  We
now make use of the evolution equation for the scalar $v =
\sqrt{v_{\alpha} v^{\alpha}}$, which can be derived from
Eq.~(\ref{dlvdotf}). The equation for $v^{2}$ is in fact given in
Paper~I; see Eq.~(146). Using the preceding asymptotic expressions
and A2, the equation for $v$ assumes the form
\be
\lb{ds_v}
\ptl_{t}v = \frac{1}{G_{-}}\,(1-v^{2})\,(3\gam-4)\,v + g \ ,
\ee
where $g = \bigO(\e^{(-1+\eps)t})$.  Equation (\ref{ds_v}) is in
fact an asymptotically autonomous DE, a scalar version of the
vector DE discussed by Horwood {\em et al\/} in
Ref.~\ct{horetal2003}. We can thus use Theorem~B.1 in
Ref.~\ct{horetal2003}, p.~15, to obtain the limiting behavior of
$v$ as $t\rightarrow\infty$. The domain $D$ for $v$ is the interval
$0 < v < 1$, and the conditions $H_{1}$ and $H_{2}$ in
Ref.~\ct{horetal2003} are satisfied. It is straightforward to
verify that the solutions of the autonomous DE
\be
\ptl_{t}v = \frac{1}{G_{-}}\,(1-v^{2})\,(3\gam-4)\,v \ ,
\ee
with initial conditions in $D$, satisfy
\be
\lb{ds_vlim}
    \lim_{t\rightarrow\infty} v = \begin{cases}
    0 & \hsp5 \text{if $1 \leq \gam < \tfrac{4}{3}$.} \\
    1 & \hsp5 \text{if $\tfrac{4}{3} < \gam < 2$.}
    \end{cases}
\ee
It then follows from Theorem~B.1 in Ref.~\ct{horetal2003} that any
solution of Eq.~(\ref{ds_v}) with initial condition in D satisfies
Eq.~(\ref{ds_vlim}).

If $1 \leq \gam < \tfrac{4}{3}$, Eq.~(\ref{ds_v}) assumes the form
of Eqs.~(\ref{lem1_1}) and~(\ref{lem1_2}) with $k = -(3\gam-4)$ and
$l = 1-\eps$. Proposition~1 then implies
\be
\lb{ds_v1}
v = \bigO(\e^{(3\gam-4+\eps)t}) \ .
\ee

If $\tfrac{4}{3} < \gam < 2$, we write the evolution equation for
$v$ [\,cf.~Paper~I, Eq.~(146)\,] in the form
\be
\lb{ds_v2}
\ptl_{t}(1-v^{2}) = - \,v^{\alpha}\,\parb_{\alpha}(1-v^{2})
- \frac{2}{G_{-}}\left[\,(3\gam-4)\,v^{2} + \bigO(\e^{(-1+\eps)t})
\,\right] (1-v^{2}) \ .
\ee
P1 and A2 imply that for any given $\delta > 0$,\footnote{Since
$1-v^{2}\rightarrow0$ in this case, we can use A2 to write
$\ptl_{i}(1-v^{2})=\bigO(1-v^{2})$, and proceed as in
footnote~\ref{foot1}.}
\be
-\,v^{\alpha}\,\parb_{\alpha}(1-v^{2}) \leq \delta \,(1-v^{2}) \ .
\ee
Then Eq.~(\ref{ds_v2}) assumes the form of Eqs.~(\ref{lem1_1}) and
(\ref{lem1_2}) with $k = 2\,\frac{(3\gam-4)}{(2-\gam)}$ and
$B_{\alpha\beta} = 0$. Proposition~1 implies
\be
\lb{ds_v3}
1-v^{2} = \bigO(\e^{[-2\frac{(3\gam-4)}{(2-\gam)}+\eps]t}) \ .
\ee

For $\gam = \tfrac{4}{3}$, Eq.~(\ref{dlvdotf}) gives
\be
\ptl_{t}v^{\alpha} = \bigO(\e^{(-1+\eps)t}) \ ,
\ee
which implies
\be
\lb{ds_v5}
v^{\alpha} = \hat{v}^{\alpha} + \bigO(\e^{(-1+\eps)t}) \ .
\ee

Using Eq.~(\ref{ds_vlim}), the inequality~(\ref{ds_Om_1}) can be
strengthened to read
\be
\lb{ds_Om_6}
\ptl_{t}\Om \leq \begin{cases}
(-3\gam+\eps)\,\Om
& \hsp5 \text{for $1 \leq \gam < \tfrac{4}{3}$,} \\
(-4+\eps)\,\Om
& \hsp5 \text{for $\tfrac{4}{3} < \gam < 2$,}
    \end{cases}
\ee
where the $\eps$ has been redefined. If $\gam = \tfrac{4}{3}$,
Eq.~(\ref{ds_Om_1}) simplifies directly to
\be
\lb{ds_Om_7}
\ptl_{t}\Om \leq (-4+\eps)\,\Om \ .
\ee
Inequalities~(\ref{ds_Om_6}) and~(\ref{ds_Om_7}) now give
\be
\Om = \begin{cases}
\bigO(\e^{(-3\gam+\eps)t})
& \hsp5 \text{for $1 \leq \gam < \tfrac{4}{3}$.} \\
\bigO(\e^{(-4+\eps)t})
& \hsp5 \text{for $\tfrac{4}{3} \leq \gam < 2$.}
    \end{cases}
\ee
%

%-----------------------
\subsubsection*{Stage 2}
%-----------------------
In the second stage we use the asymptotic decay rates found in
Stage~1 to successively write the evolution equations in a form to
which Proposition~3 can be applied.

We begin by considering $E_{\alpha}{}^{i}$.  The decay rates in
Stage~1 and Eq.~(\ref{hdecel}) gives
\be
\lb{ds_r2_q}
q = -\,1 + \bigO(\e^{(-2+\eps)t}) \ .
\ee
The evolution equation~(\ref{dl13comts}) for $E_{\alpha}{}^{i}$ can
now be written in the form
\be
\lb{ds_r2_E}
\ptl_{t}E_{\alpha}{}^{i} = -\,E_{\alpha}{}^{i} + g_{\alpha}{}^{i}
\ ,
\ee
where $g_{\alpha}{}^{i} = \bigO (\e^{(-3+\eps)t})$. This equation,
in conjunction with Proposition~3, implies that
\be
E_{\alpha}{}^{i}= \hat{E}_{\alpha}{}^{i}\,\e^{-t}
+ \bigO (\e^{(-3+\eps)t}) \ .
\ee

In a similar way the evolution equations~(\ref{dladot})
and~(\ref{dlndot}) lead to
\bea
A^{\alpha} & = & \hat{A}^{\alpha}\,\e^{-t}
+ \bigO(\e^{(-3+\eps)t}) \\
N^{\alpha\beta} & = & \hat{N}^{\alpha\beta}\,\e^{-t}
+ \bigO(\e^{(-3+\eps)t}) \ .
\eea

We now consider the evolution equation~(\ref{dlomdot}) for $\Om$.
For $1 \leq \gam < \tfrac{4}{3}$, we obtain
\be
\ptl_{t}\Om = -\,3\gam\,\Om + g \ ,
\ee
where $g = \bigO(\e^{(3\gam-8+\eps)t})$, with $\Om v^{2}$
being the dominant term in $g$. For $\tfrac{4}{3} < \gam < 2$, we
obtain
\be
\ptl_{t}\Om = -\,4\Om + g \ ,
\ee
where $g = \bigO(\e^{(-5+\eps)t} +
\e^{(-4-2\frac{(3\gam-4)}{(2-\gam)}+\eps)t})$, with
$\parb_{\alpha}\Om$, $\Om\,\parb_{\alpha}v^{\alpha}$,
$(A_{\alpha}v^{\alpha})\,\Om$ and $\Om\,(1-v^{2})$ being the
dominant terms in $g$. For $\gam = \tfrac{4}{3}$, we obtain
\be
\ptl_{t}\Om = -\,4\Om + \bigO(\e^{(-5+\eps)t}) \ .
\ee
Applying Proposition~3 yields
\be
\Om = \begin{cases}
\hat{\Om}\,\e^{-3\gam t} + \bigO(\e^{(3\gam-8+\eps)t})
& \hsp5 \text{for $1 \leq \gam < \tfrac{4}{3}$.} \\
\hat{\Om}\,\e^{-4t} + \bigO(\e^{(-5+\eps)t}
+ \e^{(-4-2\frac{(3\gam-4)}{(2-\gam)}+\eps)t})
& \hsp5 \text{for $\tfrac{4}{3} < \gam < 2$.} \\
\hat{\Om}\,\e^{-4t} +\bigO(\e^{(-5+\eps)t})
& \hsp5 \text{for $\gam =\tfrac{4}{3}$.}
\end{cases}
\ee

At this stage, the evolution equation for $\Sig^{\alpha\beta}$
assumes the form
\be
\ptl_{t}\Sig^{\alpha\beta} = -\,3\Sig^{\alpha\beta}
- 3\hat{\cal S}^{\alpha\beta}\,\e^{-2t} + g^{\alpha\beta} \ ,
\ee
where $g^{\alpha\beta}=\bigO(\e^{(-4+\eps)t})$. By imitating the
proof of Proposition~3, we conclude that
\be
\Sig^{\alpha\beta} = -\,3\hat{\cal S}^{\alpha\beta}\,\e^{-2t}
+ \hat{\Sig}^{\alpha\beta}\,\e^{-3t} + \bigO(\e^{(-4+\eps)t}) \ .
\ee
At this stage $({\cal C}_{\rm G})$ and A3 give
\be
\lb{ds_oml}
\Oml-1 = -\,\hat{\Om}_{k}\,e^{-2t}
+ \bigO(\e^{-3\gam t}+\e^{-4t}) \ ,
\ee
while $({\cal C}_{\Lambda})_{\alpha}$ and A3 lead to
\be
\lb{ds_r2_r}
r_{\alpha} = -\,\tfrac{1}{2}\,(\hat{E}_{\alpha}{}^{i}\,
\ptl_i\hat{\Om}_{k})\,\e^{-3t}
+ \bigO(\e^{-(1+3\gam)t}+\e^{-5t}) \ .
\ee

We now consider the evolution equation~(\ref{dlvdotf}) for
$v^{\alpha}$.  For $\gam = 1$, we obtain
\be
\ptl_{t}v^{\alpha} = -\,v^{\alpha} + \bigO(\e^{(-3+\eps)t}) \ .
\ee
Proposition~3 implies
\be
v^{\alpha} = \hat{v}^{\alpha}\,e^{-t} + \bigO(\e^{(-3+\eps)t}) \ .
\ee
For $1 < \gam < \tfrac{4}{3}$, we obtain
\be
\ptl_{t}v^{\alpha} = (3\gam-4)\,v^{\alpha} + g^{\alpha} \ ,
\ee
where $g^{\alpha} = \bigO(\e^{-t} + e^{3(3\gam-4+\eps)t})$, with
$\parb_{\alpha}\ln\Om$ and $v^{2}\,v^{\alpha}$ being the dominant
terms in $g^{\alpha}$. Proposition~3 implies
\be
v^{\alpha} = \hat{v}^{\alpha}\,e^{(3\gam-4)t} +
\bigO(\e^{-t} + e^{3(3\gam-4+\eps)t}) \ .
\ee
For $\gam = \tfrac{4}{3}$, we already have the result:
\be
v^{\alpha} = \hat{v}^{\alpha} + \bigO(\e^{(-1+\eps)t}) \ .
\ee
For $\tfrac{4}{3} < \gam < 2$, we obtain from Eq.~(\ref{ds_v2})
\be
\ptl_{t}(1-v^{2}) = -\,2\,\frac{(3\gam-4)}{(2-\gam)}\,(1-v^{2})
+ g \ ,
\ee
where $g = \bigO(\e^{(-1-2\frac{(3\gam-4)}{(2-\gam)}+\eps)t} +
\e^{(-4\frac{(3\gam-4)}{(2-\gam)}+\eps)t})$, with
$(1-v^{2})\,\parb_{\alpha}v^{\alpha}$,
$(1-v^{2})\,A_{\alpha}v^{\alpha}$ and $(1-v^{2})^{2}$ being the
dominant terms in $g$. Proposition~3 implies
\be
1-v^{2} = (1-\hat{v}^{2})\,\e^{-2\frac{(3\gam-4)}{(2-\gam)}t}
+ \bigO(\e^{(-1-2\frac{(3\gam-4)}{(2-\gam)}+\eps)t}
+ \e^{(-4\frac{(3\gam-4)}{(2-\gam)}+\eps)t}) \ .
\ee

Using the results obtained in Stage~2, we can repeat Stage~2 to
eliminate the epsilons in the $\bigO$-terms as follows.
Equation~(\ref{ds_r2_q}) now reads $q = -\,1 + \bigO(\e^{-2t})$,
and the function $g_{\alpha}{}^{i}$ in Eq.~(\ref{ds_r2_E})
satisfies $g_{\alpha}{}^{i} = \bigO(\e^{-3t})$; similarly for the
other variables.

The position-dependent coefficients $\hat{E}_{\alpha}{}^{i}$,
$\hat{A}^{\alpha}$ and $\hat{N}_{\alpha\beta}$ are required to
satisfy the constraints
\begin{xalignat}{2}
\lb{dS_Ccom}
0 & = 2\,(\hat{E}_{[\alpha}{}^{j}\,\ptl_{j}-\hat{A}_{[\alpha})\,
\hat{E}_{\beta]}{}^{i} -\eps_{\alpha\beta\delta}\,
\hat{N}^{\delta\gam}\,\hat{E}_{\gam}{}^{i} \\
\lb{dS_CJ}
0 & = \hat{E}_{\beta}{}^{i}\,\ptl_{i}(\hat{N}^{\alpha\beta}
+\eps^{\alpha\beta\gam}\,\hat{A}_{\gam})
- 2\hat{A}_{\beta}\,\hat{N}^{\alpha\beta} \ ,
\end{xalignat}
which arise, respectively, from $({\cal C}_{\rm
com})^{i}{}_{\alpha\beta}$ and $({\cal C}_{\rm J})^{\alpha}$, at
order $\e^{-2t}$. Equation~(\ref{dS_Ccom}) can be solved to express
the coefficients $\hat{A}^{\alpha}$ and $\hat{N}_{\alpha\beta}$ in
terms of the coefficients $\hat{E}_{\alpha}{}^{i}$ and their
inverse $\hat{E}^{\alpha}{}_{i}$. These expressions, given by
Eqs.~(\ref{Ahat}) and~(\ref{Nhat}), satisfy Eq.~(\ref{dS_CJ})
identically. The constraint $({\cal C}_{\rm C})^{\alpha}$ provides
a restriction on $\hat{v}^{\alpha}$, given by Eqs.~(\ref{dS_CC})
and~(\ref{dS_CC2}).  \hfill $\Box$

%%%%%%%%%%%%%%%%%%%%%%%%%%%%%%%%%%%%%%%%%%%%%%%%%%%%%%%%%%%%%%%%%%%
\section{Derivation of asymptotically flat-FL past dynamics}
\lb{app:flatfl}
%%%%%%%%%%%%%%%%%%%%%%%%%%%%%%%%%%%%%%%%%%%%%%%%%%%%%%%%%%%%%%%%%%%
In this appendix we derive the asymptotic expansions
(\ref{flatfl_bigO})--(\ref{flatfl_bigO_end}) for $G_{0}$
cosmologies that are past asymptotic to the flat FL solution in the
sense of Eqs.~(\ref{flatfl_c1})--(\ref{flatfl_c3}). For ease of
reference, we list the complete set of restrictions below.
\begin{enumerate}[{A}1:]
\item $E_{\alpha}{}^{i},\ A^{\alpha},\ N_{\alpha\beta},\
      r_{\alpha},\ \Sig_{\alpha\beta},\ 1-\Om,\ \Oml \rightarrow 0$
      as $t \rightarrow -\infty$, for fixed $x^{i}$.
\item The partial derivatives of any dimensionless variable $V$ are
      bounded as $t\rightarrow-\infty$. In addition, if $V$ tends to
      zero, then $\ptl_{i}V = \bigO(\| V \|)$ as
      $t\rightarrow-\infty$, for fixed $x^{i}$, where $\| V \|$ is
      the magnitude of $V$.
\item If $V = \hat{V}\,\e^{nt} + \bigO(\e^{mt})$ as $t\rightarrow
      -\infty$, where $m>n>0$, then $\ptl_{i}V =
      \ptl_{i}\hat{V}\,e^{nt} + \bigO(e^{mt})$ as $t\rightarrow
      -\infty$.
\end{enumerate}

We organize the derivation into two stages, in which we repeatedly
use the evolution equations, (\ref{ralpha}) and the constraints
$({\cal C}_{\rm G})$ and $({\cal C}_{\rm
C})^{\alpha}$. In Stage~1, we obtain order estimates for all
variables, primarily using Propositions~1 and~2 in
App.~\ref{app:prop}. In Stage 2, we use the results in Stage~1 in
conjunction with Proposition~3 to obtain the explicit asymptotic
expansions. The assumptions A2 and A3 are needed throughout in
order to bound the spatial derivative terms in the evolution
equations, in (\ref{ralpha}), and in the constraints
$({\cal C}_{\rm G})$ and~$({\cal C}_{\rm C})^{\alpha}$.

%-----------------------
\subsubsection*{Stage 1}
%-----------------------
First, A1, A2 and $({\cal C}_{\rm C})^{\alpha}$ imply that
$(\gam/G_{+})\,\Om v^{\alpha} \rightarrow 0$.
But we have $\Om \rightarrow 1$. Thus we have
\be
\lim_{t\rightarrow-\infty}v^{\alpha} = 0 \ .
\ee
It follows from A1, A2 and (\ref{hdecel}) that
\be
\lb{fl_r1_q}
\lim_{t\rightarrow-\infty}q = \tfrac{1}{2}\,(3\gam-2) \ .
\ee
We now introduce $\tau = -t$, for convenience in using
Propositions~1--3.

We begin by considering $E_{\alpha}{}^{i}$. Using A1 and
Eq.~(\ref{fl_r1_q}), the evolution equation~(\ref{dl13comts}) for
$E_{\alpha}{}^{i}$ assumes the form of Eqs.~(\ref{lem1_1}) and
(\ref{lem1_2}) with $k = \tfrac{1}{2}\,(3\gam-2)$ and $B_{\alpha\beta} =
0$. It follows from Proposition~1 that for any given
$\eps_{E} > 0$,
\be
\lb{fl_r1_E}
E_{\alpha}{}^{i}
= \bigO(\e^{[-\tfrac{1}{2}(3\gam-2)+\eps_{E}]\tau}) \ ,
\ee
as $\tau \rightarrow \infty$.

We can now use Eqs.~(\ref{fl_r1_q}), (\ref{ds_r1_E}) and A1, A2 to
conclude that the evolution equation~(\ref{dladot}) for
$A^{\alpha}$ assumes the form of Eqs.~(\ref{lem1_1}) and
(\ref{lem1_2}) with $k = \tfrac{1}{2}\,(3\gam-2)$ and $l =
\tfrac{1}{2}\,(3\gam-2)-\eps_{E}$. It follows from Proposition~1
that for any given $\eps_{A} > \eps_{E}$,
\be
\lb{fl_r1_A}
A^{\alpha} = \bigO(\e^{[-\tfrac{1}{2}(3\gam-2)+\eps_{A}]\tau}) \ ,
\ee
as $\tau \rightarrow \infty$. We make the $\eps$-simplification as
in Subsec.~\ref{app:ds}.

In a similar way, Eqs.~(\ref{dlndot}), (\ref{dlrdot}) and
(\ref{dlomldot}) lead to\footnote{Here and in the future, we will
omit the qualifier $\tau\rightarrow\infty$.}
\bea
\lb{fl_r1_N}
N^{\alpha\beta} & = & \bigO(\e^{[-\tfrac{1}{2}(3\gam-2)+\eps]\tau})
\\
\lb{fl_r1_r}
r_{\alpha} & = & \bigO(\e^{[-\tfrac{1}{2}(3\gam-2)+\eps]\tau}) \\
\lb{fl_r1_oml}
\Oml & = & \bigO(\e^{(-3\gam+\eps)\tau}) \ .
\eea

With Eqs.~(\ref{fl_r1_E})--(\ref{fl_r1_r}) and A2,
the constraint $({\cal C}_{\rm C})^{\alpha}$ gives
\be
\lb{fl_r1_v}
v^{\alpha} = \bigO(\e^{[-\tfrac{1}{2}(3\gam-2)+\eps]\tau}) \ .
\ee

Equation~(\ref{dlsigdot}) now assumes the form of
Eqs.~(\ref{lem2_1}) and (\ref{lem2_2}) with $k =
\tfrac{3}{2}\,(2-\gam)$ and $l = (3\gam-2)-\eps$. Since
$\Sig_{\alpha\beta}$ is bounded as $\tau\rightarrow\infty$ (it is
assumed to tend to zero), Proposition~2 implies
\be
\lb{fl_r1_Sig}
\Sig^{\alpha\beta} = \bigO(\e^{[-(3\gam-2)+\eps]\tau}) \ .
\ee

Lastly, the constraint $({\cal C}_{\rm G})$ implies
\be
\lb{fl_r1_Om}
1-\Om = \bigO(\e^{[-(3\gam-2)+\eps]\tau}) \ .
\ee
%

%-----------------------
\subsubsection*{Stage 2}
%-----------------------
In the second stage we use the asymptotic decay rates found in
Stage~1 to successively write the evolution equations in a form to
which Proposition~3 can be applied.

We begin by considering $E_{\alpha}{}^{i}$. The decay rates in
Stage~1 and Eq.~(\ref{hdecel}) give
\be
q = \tfrac{1}{2}\,(3\gam-2) + \bigO(\e^{[-(3\gam-2)+\eps]\tau}) \ .
\ee
The evolution equation~(\ref{dl13comts}) for $E_{\alpha}{}^{i}$ can
now be written in the form
\be
\ptl_{\tau}E_{\alpha}{}^{i} = -\,\tfrac{1}{2}\,(3\gam-2)\,
E_{\alpha}{}^{i} + g_{\alpha}{}^{i} \ ,
\ee
where $g_{\alpha}{}^{i} =
\bigO(\e^{[-\tfrac{3}{2}(3\gam-2)+\eps]\tau})$.  This equation, in
conjunction with Proposition~3, implies that
\be
E_{\alpha}{}^{i} = \hat{E}_{\alpha}{}^{i}\,
\e^{-\tfrac{1}{2}(3\gam-2)\tau}
+ \bigO(\e^{[-\tfrac{3}{2}(3\gam-2)+\eps]\tau}) \ .
\ee

In a similar way the evolution equations~(\ref{hq}),
(\ref{dlomldot}), (\ref{dladot}), (\ref{dlndot}) and (\ref{dlrdot})
lead to
\bea
H & = & \hat{H}\,\e^{(3\gam/2)\tau}
+ \bigO(\e^{[(3\gam/2)-(3\gam-2)+\eps]\tau}) \\
\Oml & = & \hat{\Om}_{\Lambda}\,\e^{-3\gam \tau}
+ \bigO(\e^{[-3\gam -(3\gam-2)+\eps]\tau}) \\
A^{\alpha} & = & \hat{A}^{\alpha}\,\e^{-\tfrac{1}{2}(3\gam-2)\tau}
+ \bigO(\e^{[-\tfrac{3}{2}(3\gam-2)+\eps]\tau}) \\
N^{\alpha\beta} & = & \hat{N}^{\alpha\beta}\,
\e^{-\tfrac{1}{2}(3\gam-2)\tau}
+ \bigO(\e^{[-\tfrac{3}{2}(3\gam-2)+\eps]\tau}) \\
r_{\alpha} & = & \hat{r}_{\alpha}\,\e^{-\tfrac{1}{2}(3\gam-2)\tau}
+ \bigO(\e^{[-\tfrac{3}{2}(3\gam-2)+\eps]\tau}) \ ,
\eea
where $\hat{\Om}_{\Lambda} = \Lambda/(3\hat{H}^{2})$, and
$\hat{r}_{\alpha} =
-\,(\hat{E}_{\alpha}{}^{i}\,\ptl_{i}\hat{H})/\hat{H}$
from (\ref{ralpha}).

At this stage, $({\cal C}_{\rm C})^{\alpha}$ and A3 give
\be
v^{\alpha} = -\,\frac{2}{3\gam}\,\hat{r}^{\alpha}\,
\e^{-\tfrac{1}{2}(3\gam-2)\tau}
+ \bigO(\e^{[-\tfrac{3}{2}(3\gam-2)+\eps]\tau}) \ ,
\ee
while $({\cal C}_{\rm G})$ and A3 lead to
\be
\Om = 1 - \hat{\Om}_{k}\, \e^{-(3\gam-2)\tau}
+ \bigO(\e^{[-2(3\gam-2)+\eps]\tau}+e^{-3\gam\tau}) \ .
\ee

We now use the freedom to re-define the 1-parameter family of
spacelike 3-surfaces ${\cal S}$:$\{t=\mbox{constant}\}$ while
preserving the separable volume gauge\footnote{A transformation of
the local coordinates of the form $\tilde{t} = t + \varphi(x^{i})+
\bigO(\e^{(3\gam-2)t})$, $\tilde{x}^{i} = x^{i} +
\frac{1}{(3\gam-2)}\,\hat{H}^{-2} \,\hat{g}^{ij} \, \ptl_j \varphi
\, \e^{(3\gam-2)t} + \bigO(\e^{2(3\gam-2)t} + \e^{3\gam t})$ will
preserve the separable volume gauge.}  to set $\hat{H} =
(2/3\gam)\,T_{0}^{-1}$, with $T_{0}$ a positive real-valued
constant. As a result,
\be
\hat{r}_{\alpha} = 0 \ ,
\ee
and $v^{\alpha} = \bigO(\e^{[-\tfrac{3}{2}(3\gam-2)+\eps]\tau})$.

The evolution equation~(\ref{dlsigdot}) for $\Sig^{\alpha\beta}$
assumes the form
\be
\lb{fl_Sig}
\ptl_{\tau}\Sig^{\alpha\beta} = \tfrac{3}{2}\,(2-\gam)\,
\Sig^{\alpha\beta} + 3\hat{\mathcal{S}}^{\alpha\beta}\,
\e^{-(3\gam-2)\tau} + g^{\alpha\beta} \ ,
\ee
where $g^{\alpha\beta}=\bigO(\e^{[-2(3\gam-2)+\eps]\tau})$. We
rewrite Eq.~(\ref{fl_Sig}) as
\be
\lb{fl_int_Sig}
\ptl_{\tau}\left(\e^{-\tfrac{3}{2}(2-\gam)\tau}\,\Sig^{\alpha\beta}
\right) = 3\hat{\mathcal{S}}^{\alpha\beta}\,
\e^{-(1+\tfrac{3}{2}\gam)\tau}
+ g^{\alpha\beta}\,\e^{-\tfrac{3}{2}(2-\gam)\tau} \ ,
\ee
and integrate from $\tau$ to $\infty$ to obtain
\be
\Sig^{\alpha\beta} = -\,\frac{6}{(3\gam+2)}\,
\hat{\mathcal{S}}^{\alpha\beta}\,\e^{-(3\gam-2)\tau}
+ \bigO(\e^{[-2(3\gam-2)+\eps]\tau}) \ .
\ee

Lastly, we derive the leading order coefficients for $r_{\alpha}$
and $v^{\alpha}$. First,
\be
q = \tfrac{1}{2}\,(3\gam-2) - \tfrac{1}{2}\,(3\gam-2)\,\hat{\Om}_{k}
\,\e^{-(3\gam-2)\tau}
+ \bigO(\e^{[-2(3\gam-2)+\eps]\tau}+\e^{-3\gam\tau}) \ .
\ee
Evolution equation~(\ref{hq}) for $H$ then yields
\be
H = \hat{H}\,\e^{(3\gam/2)\tau}
\left[\,1 + \tfrac{1}{2}\,\hat{\Om}_{k}\,\e^{-(3\gam-2)\tau}
+ \bigO(\e^{[-2(3\gam-2)+\eps]\tau}+\e^{-3\gam \tau})\,\right]  \ .
\ee
Eq.~(\ref{ralpha}) and the constraint $({\cal C}_{\rm C})^{\alpha}$ then
successively give
\be
r_{\alpha} = \e^{-\tfrac{3}{2}(3\gam-2)\tau}
\left[\,-\,\tfrac{1}{2}\,(\hat{E}_{\alpha}{}^{i}\,
\ptl_{i}\hat{\Om}_{k})
+ \bigO(\e^{[-2(3\gam-2)+\eps]\tau} +\e^{-3\gam \tau}) \right] \ ,
\ee
and
\be
v^{\alpha} = \e^{-\tfrac{3}{2}(3\gam-2)\tau}
\left[\,\hat{v}^{\alpha}
+ \bigO(\e^{[-2(3\gam-2)+\eps]\tau} +\e^{-3\gam \tau})\,\right] \ ,
\ee
where
\be
\lb{vhat}
\hat{v}^{\alpha} = \frac{1}{3\gam}\,\d^{\alpha\beta}\,
(\hat{E}_\beta{}^{i}\,\ptl_{i}\hat{\Om}_{k})
+ \frac{2}{\gam(3\gam+2)}\left(\hat{E}_\beta{}^{i}\,
\ptl_{i}\hat{\mathcal{S}}^{\alpha\beta}
-3\hat{A}_{\beta}\,\hat{\mathcal{S}}^{\alpha\beta}
- \eps^{\alpha\beta\gam}\,\hat{N}_{\beta\delta}\,
\hat{S}_{\gam}{}^{\delta}\right) \ .
\ee
The twice-contracted 3-Bianchi identity~(\ref{dl3bianid}) gives
\be
0 = \hat{E}_{\beta}{}^{i}\,\ptl_{i}\hat{\mathcal{S}}^{\alpha\beta}
- 3\hat{A}_{\beta}\,\hat{\mathcal{S}}^{\alpha\beta}
- \eps^{\alpha\beta\gam}\,\hat{N}_{\beta\delta}\,
\hat{S}_{\gam}{}^{\delta}
+ \tfrac{1}{3}\, \d^{\alpha\beta}\,(\hat{E}_\beta{}^{i}\,
\ptl_{i}\hat{\Om}_{k}) \ .
\ee
This simplifies $\hat{v}^{\alpha}$ to
\be
\hat{v}^{\alpha} = \frac{1}{(3\gam+2)}\,
\d^{\alpha\beta}\,(\hat{E}_{\beta}{}^{i}\,\ptl_{i}\hat{\Om}_{k}) \ .
\ee

Using the results obtained in Stage~2, we can repeat Stage~2 to
eliminate the epsilons in the $\bigO$-terms, as in
App.~\ref{app:ds}.

As in App.~\ref{app:ds}, the position-dependent coefficients
$\hat{E}_{\alpha}{}^{i}$, $\hat{A}^{\alpha}$ and
$\hat{N}_{\alpha\beta}$ are required to satisfy the
constraints~(\ref{dS_Ccom}) and~(\ref{dS_CJ}), which arise,
respectively, from $({\cal C}_{\rm com})^{i}{}_{\alpha\beta}$ and
$({\cal C}_{\rm J})^{\alpha}$, at order $\e^{-(3\gam-2)\tau}$.
Equation~(\ref{dS_Ccom}) can be solved to express the coefficients
$\hat{A}^{\alpha}$ and $\hat{N}_{\alpha\beta}$ in terms of the
coefficients $\hat{E}_{\alpha}{}^{i}$ and their inverse
$\hat{E}^{\alpha}{}_{i}$. These expressions, given by
Eqs.~(\ref{Ahat}) and~(\ref{Nhat}), satisfy Eq.~(\ref{dS_CJ})
identically.  \hfill $\Box$

%%%%%%%%%%%%%%%%%%%%%%%%%%%%%%%%%%%%%%%%%%%%%%%%%%%%%%%%%%%%%%%%%%%
\section{Asymptotic expansions for event horizons}
\lb{sec:eh_ae}
%%%%%%%%%%%%%%%%%%%%%%%%%%%%%%%%%%%%%%%%%%%%%%%%%%%%%%%%%%%%%%%%%%%
Equation~(\ref{tadef}) and expansion~(\ref{dS_bigO}) imply that
$t_{\alpha} = \bigO(\e^{-t})$ as $t\rightarrow\infty$, subject to the
assumption that $\mathcal{E}$ satisfies A2.
Equations~(\ref{sexpr}) and~(\ref{sdef}) successively imply that
\be
s = \bigO(\e^{-t}) \ , \hspace{10mm}
\mathcal{E} = \bigO(\e^{-t}) \ .
\ee
With these results, the evolution equation~(\ref{Keq}) for
$K^{\alpha}$ yields $\ptl_{t}K^{\alpha} = \bigO(\e^{-t})$, which
can be integrated to give
\be
\lb{K_ae}
K^{\alpha} = \hat{K}^{\alpha} + \bigO(\e^{-t}) \ .
\ee
With expansions~(\ref{dS_bigO}) and (\ref{K_ae}), Eq.~(\ref{geo}) is
integrated to give
\be
x^{i}(t) = x^{i}_{\infty} + \bigO(\e^{-t})\ .
\ee
The coefficients $\hat{E}_{\alpha}{}^{i}[x^{j}(t)]$ are then
Taylor-expanded:
\be
\lb{Ex_ae}
\hat{E}_{\alpha}{}^{i}[x^{j}(t)]
= \hat{E}_{\alpha}{}^{i}(x^{j}_{\infty}) + \bigO(\e^{-t}) \ .
\ee
With expansions~(\ref{dS_bigO}), (\ref{K_ae}) and (\ref{Ex_ae}),
Eq.~(\ref{geo}) is integrated to give an improved expansion
\be
\lb{event_ds2}
x^{i}(t) = x^{i}_{\infty} - \hat{K}^{\alpha}
\hat{E}_{\alpha}{}^{i}(x^{j}_{\infty})\,\e^{-t}
+ \bigO(\e^{-2t}) \ .
\ee
%

%%%%%%%%%%%%%%%%%%%%%%%%%%%%%%%%%%%%%%%%%%%%%%%%%%%%%%%%%%%%%%%%%%%
\section{Asymptotic expansions for particle horizons}
\lb{sec:ph_ae}
%%%%%%%%%%%%%%%%%%%%%%%%%%%%%%%%%%%%%%%%%%%%%%%%%%%%%%%%%%%%%%%%%%%
Equation (\ref{tadef}) and expansion (\ref{flatfl_bigO}) imply that
$t_{\alpha} = \bigO(\Eta)$ as $\Eta\rightarrow 0$, subject to the
assumption that $1/\mathcal{E}$ satisfies
A2. Equations~(\ref{sexpr}) and~(\ref{sdef}) successively imply
that
\be
s = \bigO(\Eta) \ , \hspace{10mm}
1/\mathcal{E} = \bigO(\Eta^{2/(3\gam-2)}) \ .
\ee
With these results the evolution equation~(\ref{Keq}) for
$K^{\alpha}$ yields
\be
\lb{fl_K_ae}
K^{\alpha} = \hat{K}^{\alpha} + \bigO(\Eta) \ .
\ee
With expansions~(\ref{flatfl_bigO}) and (\ref{fl_K_ae}), Eq.~(\ref{geo})
is integrated to give
\be
x^{i}(\Eta) = x^{i}_{\rm BB} + \bigO(\Eta^{2}) \ .
\ee
The coefficients $\hat{E}_{\alpha}{}^{i}[x^{j}(\Eta)]$ are then
Taylor-expanded:
\be
\lb{fl_Ex_ae}
\hat{E}_{\alpha}{}^{i}[x^{j}(\Eta)]
= \hat{E}_{\alpha}{}^{i}(x^{j}_{\rm BB}) + \bigO(\Eta) \ .
\ee
With expansions~(\ref{flatfl_bigO}), (\ref{fl_K_ae}) and
(\ref{fl_Ex_ae}), Eq.~(\ref{geo}) is integrated to give an improved
expansion
\be
\lb{par_fl2}
x^{i}(\Eta) = x^{i}_{\rm BB} + \frac{2}{(3\gam-2)}\,
\hat{K}^{\alpha}\hat{E}_{\alpha}{}^{i}(x^{j}_{\rm BB})\,\Eta
+ \bigO(\Eta^{2}) \ .
\ee
%

%%%%%%%%%%%%%%%%%%%%%%%%%%%%%%%%%%%%%%%%%%%%%%%%%%%%%%%%%%%%%%%%%%%
\section{Fluid kinematical variables via boost transformations}
\lb{subsec:boost}
%%%%%%%%%%%%%%%%%%%%%%%%%%%%%%%%%%%%%%%%%%%%%%%%%%%%%%%%%%%%%%%%%%%
Let $\ti{\vec{u}}$ denote a fluid 4-velocity vector field. Then
fluid kinematical variables arise in the irreducible decomposition
of the spacetime gradient of $\ti{\vec{u}}$, i.e.,
\be
\nabla_{a}\ti{u}_{b} = -\,\ti{u}_{a}\,\dot{\ti{u}}_{b}
+ \ti{H}\,\ti{h}_{ab} + \ti{\sig}_{ab}
+ \ti{\om}_{ab} \ ,
\ee
defining the fluid acceleration $\dot{\ti{u}}{}^{a}$, the fluid
Hubble scalar $\ti{H}$, the fluid shear rate $\ti{\sig}_{ab}$, and
the fluid vorticity $\ti{\om}_{ab}$, respectively.

With respect to an Eulerian reference frame, comoving with a
unit timelike congruence $\vec{u}$, we have
\be
\ti{u}{}^{a} = \Gam\,(u^{a}+v^{a}) \ , \hspace{10mm}
\ti{h}{}^{a}{}_{b} = h^{a}{}_{b}
+ \Gam^{2}\,(v^{2}u^{a}+v^{a})\,u_{b}
+ \Gam^{2}\,(u^{a}+v^{a})\,v_{b} \ ,
\ee
where $h^{a}{}_{b} := \d^{a}{}_{b} + u^{a}u_{b}$, and the Lorentz
factor is defined by $\Gam := 1/\sqrt{1-v^{2}}$, with $v^{2} :=
v_{a}v^{a}$. We define the peculiar kinematical variables
\be
\lb{notat}
\hat{\th}(\vec{v}) := \D_{a}v^{a} \hspace{10mm}
\hat{\sig}_{ab}(\vec{v}) := \D_{\la a}v_{b\ra} \hspace{10mm}
\hat{\om}{}^{a}(\vec{v}) := \eps^{abc}\,\D_{b}v_{c} \ .
\ee
In the case when $\vec{u}$ is vorticity-free [\,i.e.,
$\om^{a}(\vec{u}) \equiv 0$\,], we find the boost transformations:
\enl

\noindent
{\em Fluid acceleration\/}:
\bea
\lb{flaccs}
h^{a}{}_{b}\,\dot{\ti{u}}{}^{b}
& = & \Gam^{2}\,[\,\udot^{a}+\dot{v}^{\la a\ra}
+(H+\tfrac{1}{3}\,\hat{\th})\,v^{a}
+ (\sig^{a}{}_{b}+\hat{\sig}^{a}{}_{b})\,v^{b}
+ \eps^{abc}\,\hat{\om}_{b}\,v_{c}\,] \nonumber \\
& & \hsp5 + \ \Gam^{4}\,(\dot{v}_{b}v^{b}
+\tfrac{1}{3}\,\hat{\th}\,v^{2}
+\hat{\sig}_{bc}v^{b}v^{c})\,v^{a} \\
u_{a}\dot{\ti{u}}{}^{a}
& = & -\,\dot{\ti{u}}_{a}v^{a} \ .
\eea
{\em Fluid Hubble scalar\/}:
\bea
\lb{flexp}
\ti{H} & = & \Gam\,(H+\tfrac{1}{3}\,\hat{\th}+\tfrac{1}{3}\,
\udot_{a}v^{a}) + \tfrac{1}{3}\,\Gam^{3}\,(\dot{v}_{a}v^{a}
+\tfrac{1}{3}\,\hat{\th}\,v^{2} + \hat{\sig}_{ab}v^{a}v^{b}) \ .
\eea
{\em Fluid shear rate\/}:
\bea
\lb{flshss}
h^{c}{}_{a}\,h^{d}{}_{b}\,\ti{\sig}_{cd}
& = & \Gam\,(\sig_{ab}+\hat{\sig}_{ab})
+ \Gam^{3}\,v_{(a}\,(\udot_{b)}+\dot{v}_{\la b\ra)}
+\sig_{b)c}v^{c}) + 2\Gam^{3}\,v_{(a}\,\hat{\sig}_{b)c}\,v^{c}
\nonumber \\
& & \hsp5 - \ \tfrac{1}{3}\,\Gam\,(\udot_{c}v^{c})
\,[\,h_{ab} + \Gam^{2}\,v_{a}v_{b}\,]
- \tfrac{1}{3}\,\Gam^{3}\,(\dot{v}_{c}v^{c})
\,[\,h_{ab} - 2\Gam^{2}\,v_{a}v_{b}\,] \nonumber \\
& & \hsp5 + \ \tfrac{1}{3}\,\Gam^{5}\,\hat{\th}\,v_{a}v_{b}
- \tfrac{1}{9}\,\Gam^{3}\,\hat{\th}\,v^{2}
\,[\,h_{ab} + \Gam^{2}\,v_{a}v_{b}\,]
- \tfrac{1}{3}\,\Gam^{3}\,(\hat{\sig}_{cd}v^{c}v^{d})
\,[\,h_{ab} -  2\Gam^{2}\,v_{a}v_{b}\,] \\
u^{a}u^{b}\,\ti{\sig}_{ab}
& = & \ti{\sig}_{ab}\,v^{a}v^{b} \\
u^{b}\,h^{c}{}_{a}\,\ti{\sig}_{bc}
& = & -\,\ti{\sig}_{bc}\,v^{b}\,h^{c}{}_{a} \ .
\eea
{\em Fluid vorticity\/}:
\bea
\lb{flvorss}
h^{c}{}_{a}\,h^{d}{}_{b}\,\ti{\om}_{cd}
& = & \Gam\,\eps_{abc}\,\hat{\om}^{c}
+ \Gam^{3}\,v_{[a}\,(\udot_{b]}+\dot{v}_{\la b\ra]}
+\sig_{b]c}v^{c})
+ 2\Gam^{3}\,v_{[a}\,\eps_{b]cd}\,\hat{\om}^{c}\,v^{d} \\
\lb{flvorst}
u^{b}\,h^{c}{}_{a}\,\ti{\om}_{bc}
& = & -\,\ti{\om}_{bc}\,v^{b}\,h^{c}{}_{a} \ .
\eea
Equations~(\ref{flaccs}), (\ref{flexp}), (\ref{flshss}) and
(\ref{flvorss}) provide the generalizations to $G_{0}$ cosmologies
of Eqs.~(1.15), (1.16), (1.27) and~(1.26) given by King and
Ellis~\ct{kinell73} for the SH subcase.

The covariant derivatives of $\vec{v}$ that appear in
Eqs.~(\ref{flaccs})--(\ref{flvorst}) convert into orthonormal frame
expressions according to
\bea
\dot{v}^{\la a\ra} & \longrightarrow &
\vece_{0}(v^{\alpha}) - \eps^{\alpha\beta\gam}\,
\Om_{\beta}\,v_{\gam} \\
\D_{a}v^{a} & \longrightarrow &
(\vece_{\alpha}-2a_{\alpha})\,(v^{\alpha}) \\
\D_{\la a}v_{b\ra} & \longrightarrow &
(\vece_{\la\alpha}+a_{\la\alpha})\,(v_{\beta\ra})
- \eps_{\gam\delta\la\alpha}\,n_{\beta\ra}{}^{\gam}\,v^{\delta} \\
\eps^{abc}\,\D_{b}v_{c} & \longrightarrow &
\eps^{\alpha\beta\gam}\,(\vece_{\beta}-a_{\beta})\,(v_{\gam})
- n^{\alpha}{}_{\beta}\,v^{\beta} \ .
\eea

Finally, we define Hubble-normalized fluid kinematical scalars by
\bea
\lb{fldlaccsc}
\dot{\ti{U}}{}^{2} & := &
\frac{1}{3}\left(\frac{\dot{\ti{u}}_{a}}{\ti{H}}\right)
\left(\frac{\dot{\ti{u}}{}^{a}}{\ti{H}}\right)
\ = \ \tfrac{1}{3}\,(\dot{\ti{U}}_{a}\dot{\ti{U}}{}^{a}) \\
\lb{fldlshsc}
\ti{\Sig}^{2} & := &
\frac{1}{6}\left(\frac{\ti{\sig}_{ab}}{\ti{H}}\right)
\left(\frac{\ti{\sig}^{ab}}{\ti{H}}\right)
\ = \ \tfrac{1}{6}\,(\ti{\Sig}_{ab}\ti{\Sig}^{ab}) \\
\lb{fldlvorsc}
\ti{W}^{2} & := &
\frac{1}{6}\left(\frac{\ti{\om}_{ab}}{\ti{H}}\right)
\left(\frac{\ti{\om}^{ab}}{\ti{H}}\right)
\ = \ \tfrac{1}{6}\,(\ti{W}_{ab}\ti{W}^{ab}) \ .
\eea
%

%%%%%%%%%%%%%%%%%%%%%%%%%%%%%%%%%%%%%%%%%%%%%%%%%%%%%%%%%%%%%%%%%%%

%%%%%%%%%%%%%%%%%%%%%%%%%%%%%%%%%%%%%%%%%%%%%%%%%%%%%%%%%%%%%%%%%%%

%%%%%%%%%%%%%%%%%%%%%%%%%%%%%%%%%%%%%%%%%%%%%%%%%%%%%%%%%%%%%%%%%%%
\end{document}